\documentstyle[preprint,revtex,eqsecnum]{aps}
\tightenlines
\begin{document}
\noindent{Submitted to {\it Phys.\ Rev.\ D} \hfill DOE/ER/40322-151}

\noindent{\hfill U. of Md. PP \#92-188}

\noindent{\hfill U. of Ky. PP \#UK/92-4}

\noindent{\hfill TRIUMF PP \#TRI-PP-92-28}
\draft

\vspace{18pt}

\begin{title}
{\Large Decuplet Baryon Structure from Lattice QCD}
\end{title}

\vspace{6pt}

\author{Derek B. Leinweber}

\begin{instit}
Department of Physics and Center for Theoretical Physics \\
University of Maryland, College Park, MD 20742 \cite{addr} \\
and \\
TRIUMF, 4004 Wesbrook Mall, Vancouver, B.C. Canada V6T 2A3
\end{instit}

\author{Terrence Draper}

\begin{instit}
Department of Physics and Astronomy,
University of Kentucky, Lexington, KY 40506
\end{instit}

\author{R.M. Woloshyn}

\begin{instit}
TRIUMF, 4004 Wesbrook Mall, Vancouver, B.C. Canada V6T 2A3
\end{instit}

\begin{abstract}
   The electromagnetic properties of the $SU(3)$-flavor baryon
decuplet are examined within a lattice simulation of quenched QCD.
Electric charge radii, magnetic moments, and magnetic radii are
extracted from the $E0$ and $M1$ form factors.  Preliminary results
for the $E2$ and $M3$ moments are presented giving the first model
independent insight to the shape of the quark distribution in the
baryon ground state.  As in our octet baryon analysis, the lattice
results give evidence of spin-dependent forces and mass effects in the
electromagnetic properties.  The quark charge distribution radii
indicate these effects act in opposing directions.  Some baryon
dependence of the effective quark magnetic moments is seen.  However,
this dependence in decuplet baryons is more subtle than that for octet
baryons.  Of particular interest are the lattice predictions for the
magnetic moments of $\Omega^-$ and $\Delta^{++}$ for which new recent
experimental measurements are available.  The lattice prediction of
the $\Delta^{++}/p$ ratio appears larger than the experimental ratio,
while the lattice prediction for the $\Omega^-/p$ magnetic moment
ratio is in good agreement with the experimental ratio.
\end{abstract}

\newpage

\narrowtext

\section{INTRODUCTION}

   Knowledge of the quark substructure of baryons is largely based on
experiment and model dependent descriptions of quark-gluon
interactions.  While most models are ``QCD inspired'', they all suffer
{}from an approximate treatment of the complex nonperturbative long
distance interactions.  Often these interactions are approximated by a
simple potential.  Quark spin dependent interactions are approximated
by a single gluon or effective pion exchange.  In many models the
subtleties of current quark and gluon interactions are lumped into an
effective or constituent quark mass.  Skyrme models offer an
alternative description of hadronic phenomena.  However, the
foundation of the model is in the $1/N_c$ expansion of QCD.  To learn
the true nature of nonperturbative QCD and, we hope, hadronic
phenomena, it is necessary to calculate directly with the QCD
Lagrangian in a manner which fully accounts for nonperturbative
interactions.  The most successful, reliable and promising approach
currently available is that of numerical simulations of QCD.

   Through studies of hadronic electromagnetic form factors, the
lattice gauge approach to QCD has proved to be a valuable tool in
revealing the underlying quark substructure of hadrons
\cite{leinweber91,draper90}.  Early calculations focused on the pion
electric form factor with $SU(2)$ color
\cite{wilcox85,woloshyn86a,woloshyn86b} and later with $SU(3)$ color
\cite{draper88,martinelli88,draper89}. Calculations of the proton
electric form factor followed \cite{martinelli89}.  Electromagnetic
form factors of $\pi$, $\rho$ and $N$ were calculated
\cite{draper90} from which magnetic moments and electric charge radii
were extracted.  Our analysis of the entire baryon flavor-octet
followed \cite{leinweber91} in which electromagnetic properties were
reported for both baryons and the quark sector contributions.  This
examination of octet baryons exposed the presence of spin-dependent
forces and center-of-mass effects in the underlying quark dynamics.
These effects give rise to large variations in the quark contributions
to baryon magnetic moments which were not anticipated by model
calculations.  Recently the $q^2$ dependence of the nucleon
electromagnetic form factors was examined using a method which
characterizes one of the nucleon interpolating fields as a zero
momentum secondary source \cite{wilcox92}.

   Other studies of hadron structure in lattice QCD have been
pursued through an examination of current overlap distribution
functions \cite{wilcox86,wilcox91,chu91} and Bethe-Salpeter amplitudes
\cite{chu91}.  Form factors have a number of advantages over
Bethe-Salpeter amplitudes, in that they are gauge invariant, path
independent and allow the extraction of quark distribution radii
relative to the system center of mass.  In contrast, Bethe-Salpeter
amplitudes only give relative quark separations and subtle dynamical
effects can remain hidden.

   In this paper we continue our study of hadron structure and present
the first lattice QCD calculation of the electromagnetic form factors
of $SU(3)$-flavor-decuplet baryons.  An analysis of electromagnetic
transition moments will follow in a subsequent paper
\cite{leinweber92d}.  The technique for extracting the four form
factors associated with spin-3/2 baryons from the electromagnetic
current matrix elements has been outlined in Ref.\ \cite{nozawa90}.
{}From these form factors we will determine magnetic moments, electric
radii and magnetic radii and present preliminary results for the
higher-order $E2$ and $M3$ moments.

   On the lattice, decuplet baryons are stable as a result of the
unphysically large quark masses that are used in present calculations.
Decay to a pion and an octet baryon is forbidden by energy
conservation.  However, stability of decuplet baryons is common to
most hadronic models.  In this sense, lattice results provide a new
forum in which the strengths and weaknesses of various models may be
identified.  The lattice results also provide access to observables
not readily available with present experiments such as the
higher-order multipole moments of the $\Omega^-$ which is stable to
strong interactions.

   An examination of decuplet baryon structure in lattice QCD enables
one to study new aspects of nonperturbative quark-gluon dynamics.  In
analyzing the results we make comparisons within the baryon decuplet
and with the octet results which provide insight into the spin
dependence of quark interactions.  The $E2$ and $M3$ moments
accessible in spin-3/2 systems provide a preliminary glimpse at the
shape of the decuplet baryon ground state.  These higher-order moments
also have the potential to discriminate between model dependent
descriptions of hadronic phenomena.  For example, a vanishing $E2$
moment would cast serious doubt on hedgehog Skyrmion descriptions of
baryons.  To put our results into perspective, we compare our
calculations with experimental measurements where available, with
recent quark, bag and Skyrme model calculations, and with QCD sum rule
calculations.

   The format of our paper is as follows.  Interpolating fields are
discussed in section II A.  Correlation functions at the quark level
are discussed in section II B.  Two-point and three-point correlation
functions at the hadronic level are discussed in sections II C and II
D respectively.  Lattice techniques are discussed briefly in section
II E.  Decuplet baryon masses are reported in section III A.
Correlation function ratios used in extracting the multipole form
factors are illustrated in section III B.  Our findings for the
electromagnetic properties of the four multipole form factors are
presented in sections III C through III F.  Finally, section IV
provides an overview of our results and a discussion of future
investigations.

\section{THEORETICAL FORMALISM}

\subsection{Interpolating fields}

   The commonly used interpolating field for exciting the
$\Delta^{++}$ resonance from the QCD vacuum takes the long established
\cite{ioffe81,chung82} form of
\begin{equation}
\chi_\mu^{\Delta^{++}}(x) =
\epsilon^{abc} \left ( u^{Ta}(x) C \gamma_\mu
                                u^b(x) \right ) u^c(x).
\end{equation}
Unless otherwise noted, we follow the notation of Sakurai
\cite{sakurai67}.  The Dirac gamma matrices are Hermitian and satisfy
$\left \{ \gamma_\mu , \gamma_\nu \right \} = 2 \, \delta_{\mu
\nu}$, with $\sigma_{\mu \nu} = {1 \over 2i} \left [ \gamma_\mu ,
\gamma_\nu \right ] $.  $C = \gamma_4 \gamma_2$ is the charge
conjugation matrix, $a,\ b,\ c$ are color indices, $u(x)$ is a
$u$-quark field, and the superscript $T$ denotes transpose.  The
generalization of this interpolating field for the $\Delta^+$ composed
of two $u$ quarks and one $d$ quark has the form
\FL
\begin{eqnarray}
\chi_\mu^{\Delta^{+}}(x) =
{1 \over \sqrt{3} } \; \epsilon^{abc} \Bigl [
&2& \left ( u^{Ta}(x) C \gamma_\mu d^b(x) \right )
u^c(x) \nonumber \\
&+& \left ( u^{Ta}(x) C \gamma_\mu u^b(x) \right )
d^c(x)\ \Bigr ] \, .
\label{deltapif}
\end{eqnarray}
Other decuplet baryon interpolating fields are obtained with the
appropriate substitutions of $u(x),\ d(x)\ \to\ u(x),\ d(x)$ or
$s(x)$.  The interpolating field for $\Sigma^{*0}$ is given by the
symmetric generalization
\FL
\begin{eqnarray}
\chi_\mu^{\Sigma^{*0}}(x) =
\sqrt{2 \over 3} \; \epsilon^{abc} \Bigl [
& & \left ( u^{Ta}(x) C \gamma_\mu d^b(x) \right )
s^c(x) \nonumber \\
&+& \left ( d^{Ta}(x) C \gamma_\mu s^b(x) \right )
u^c(x) \nonumber \\
&+& \left ( s^{Ta}(x) C \gamma_\mu u^b(x) \right )
d^c(x)\ \Bigr ] \, .
\end{eqnarray}
The $SU(2)$-isospin symmetry relationship for $\Sigma^*$ form factors
\begin{equation}
\Sigma^{*0} = {\Sigma^{*+} + \Sigma^{*-} \over 2} \, ,
\end{equation}
may be easily seen in the $\Sigma^{*0}$ interpolating field by noting
\begin{eqnarray}
\lefteqn{\epsilon^{abc} \left ( s^{Ta}(x) C \gamma_\mu u^b(x) \right )
d^c(x) =} \qquad \qquad \qquad \nonumber \\
&& \epsilon^{abc}
\left ( u^{Ta}(x) C \gamma_\mu s^b(x) \right ) d^c(x) .
\end{eqnarray}

\subsection{Correlation functions}

   Two-point correlation functions at the quark level are obtained
through the standard procedure of contracting out pairs of quark
fields.  Consider the $\Delta^+$ correlation function at the quark
level.
\widetext
\begin{eqnarray}
\left < T \left ( \chi_\mu^{\Delta^+}(x),
\overline \chi_\nu^{\Delta^+}(0) \right ) \right > &=&
{1 \over 3} \; \epsilon^{abc} \epsilon^{a'b'c'}  \Bigl \{ \nonumber \\
&& \;\;\;\: 4 S_u^{a a'} \,
                       \gamma_\nu \, C S_u^{T b b'} C \, \gamma_\mu \,
S_d^{c c'} \nonumber \\
&& + \; 4 S_u^{a a'}\, \gamma_\nu \, C S_d^{T b b'} C \, \gamma_\mu \,
S_u^{c c'} \nonumber \\
&& + \; 4 S_d^{a a'}\, \gamma_\nu \, C S_u^{T b b'} C \, \gamma_\mu \,
S_u^{c c'} \label{deltapcf} \\
&& + \; 2 S_u^{a a'} \, tr \left [ \gamma_\nu \,
  C S_u^{T b b'} C \, \gamma_\mu \, S_d^{c c'} \right ] \nonumber \\
&& + \; 2 S_u^{a a'} \, tr \left [ \gamma_\nu \,
  C S_d^{T b b'} C \, \gamma_\mu \, S_u^{c c'} \right ] \nonumber \\
&& + \; 2 S_d^{a a'} \, tr \left [ \gamma_\nu \,
  C S_u^{T b b'} C \, \gamma_\mu \, S_u^{c c'} \right ] \Bigr \} ,
\nonumber
\end{eqnarray}
\narrowtext
where the quark-propagator $S_u^{a a'} = T \left ( u^a(x), \overline
u^{a'}(0) \right )$ and similarly for other quark flavors.
$SU(3)$-flavor symmetry is clearly displayed in this equation.

   The corresponding connected three-point function may be constructed
by replacing each of the three propagators $S$, one at a time, by
$\widehat S$ denoting the propagation of a quark in the presence of
the electromagnetic current.  The interesting result obtained from
(\ref{deltapcf}) is that the electromagnetic form factors of the
neutral charge decuplet baryons vanish under $SU(3)$-flavor symmetry.
Introduction of the more massive strange quark, as in $\Sigma^{*0}$ or
$\Xi^{*0}$, violates $SU(3)$-flavor symmetry and allows a nontrivial
result.  However, under $SU(2)$-isospin symmetry the electromagnetic
form factors of $\Delta^0$ vanish.  A non-zero value for the
magnetic moment of the $\Delta^0$ resonance reflects differences in
the $u$- and $d$-quark masses and contributions from the quark sea
through disconnected quark loops interacting with the electromagnetic
current.

\subsection{Two-point Green functions}

   In this and the following subsection discussing correlation
functions at the hadronic level, the Dirac representation of the
$\gamma$-matrices as defined in Itzykson and Zuber \cite{itzykson80}
is used to facilitate calculations of the $\gamma$-matrix algebra.  It
is then a simple task to account for the differences in
$\gamma$-matrix and metric definitions in reporting the final results
using Sakurai's notation.

   The extraction of baryon mass and electromagnetic form factors
proceeds through the calculation of the ensemble average (denoted
$\bigm < \cdots \bigm >$) of two and three-point Green functions. The
two-point function is defined as
\FL
\begin{eqnarray}
\lefteqn{\bigm < G^{BB}_{\sigma \tau} (t;\vec{p}; \Gamma) \bigm > = }
\qquad \nonumber \\
&& \sum_{\vec{x}} e^{-i
\vec{p} \cdot \vec{x} } \Gamma^{\beta \alpha} \bigm < \Omega \bigm |
T \left ( \chi^\alpha_{\sigma}(x)
\overline \chi^\beta_{\tau}(0) \right )
\bigm | \Omega \bigm > .
\end{eqnarray}
Here $\Omega$ represents the QCD vacuum, $\Gamma$ is a $4 \times 4$
matrix in Dirac space and $\alpha, \, \beta$ are Dirac indices.  The
subscripts $\sigma, \, \tau$ are the Lorentz indices of the spin-3/2
interpolating fields.  At the hadronic level one proceeds by inserting
a complete set of states\break $\bigm | B, p, s \bigm >$ and defining
\begin{equation}
\bigm < \Omega \bigm | \chi_{\sigma} (0) \bigm | B, p, s \bigm > \, =
   Z_B \sqrt{M \over E_p} \, u_{\sigma}(p,s) ,
\label{interp}
\end{equation}
where $Z_B$ represents the coupling strength of $\chi(0)$ to baryon
$B$.  Momentum is denoted by $p$, spin by $s$, and $u_\alpha(p,s)$ is
a spin-vector in the Rarita-Schwinger formalism \cite{rarita41}.  $E_p
= \sqrt{\vec{p}^2 + M^2}$ and Dirac indices have been suppressed.
Using the Rarita-Schwinger spin sum,
\widetext
\begin{eqnarray}
\sum_s u_\sigma(p,s) \overline u_\tau(p,s) &=&
- { \gamma \cdot p + M \over 2 M } \left \{ g_{\sigma \tau} - { 1
\over 3} \gamma_\sigma \gamma_\tau - { 2 p_\sigma p_\tau \over 3 M^2 }
+ { p_\sigma \gamma_\tau - p_\tau \gamma_\sigma \over 3 M} \right \} ,
\label{rsss} \\
&\equiv& \Lambda_{\sigma \tau} , \nonumber
\end{eqnarray}
\narrowtext
our usual definitions for $\Gamma$,
\begin{equation}
\Gamma_j = {1 \over 2} \left ( \begin{array}{cc} \sigma_j &\quad 0 \\
                                                 0        &\quad 0 \\
                               \end{array} \right ) \quad ;
\quad \Gamma_4 = {1 \over 2} \left ( \begin{array}{cc} I &\quad 0 \\
                                                       0 &\quad 0 \\
                                     \end{array} \right ) ,
\label{gammadef}
\end{equation}
and $\vec p = (p,0,0)$, the large Euclidean time limit of the two
point function takes the form
\begin{equation}
   \bigm < G^{BB}_{\sigma \tau} (t;\vec{p}, \Gamma_4) \bigm > \,
   = Z_B^2 {M \over E_p}
   e^{- E_p t} \, tr \left [\; \Gamma_4 \; \Lambda_{\sigma \tau} \;
   \right ] ,
\end{equation}
where
\narrowtext
\begin{mathletters}
\FL
\begin{eqnarray}
   \bigm < G^{BB}_{00} (t;\vec{p}, \Gamma_4) \bigm >
   &=& Z_B^2 \,
   {2 \over 3} \, { |\vec p|^2 \over M_B^2 }
   \left ( {E_p + M_B \over 2 E_p} \right ) e^{- E_p t} ,
   \label{gbb00} \\
   \bigm < G^{BB}_{11} (t;\vec{p}, \Gamma_4) \bigm >
   &=& Z_B^2 \,
   {2 \over 3} \, { E_p^2 \over M_B^2 }
   \left ( {E_p + M_B \over 2 E_p} \right ) e^{- E_p t} ,
   \label{gbb11} \\
   \bigm < G^{BB}_{22} (t;\vec{p}, \Gamma_4) \bigm >
   &=& Z_B^2 \,
   {2 \over 3}
   \left ( {E_p + M_B \over 2 E_p} \right ) e^{- E_p t} ,
   \label{gbb22} \\
   \bigm < G^{BB}_{33} (t;\vec{p}, \Gamma_4) \bigm >
   &=& Z_B^2 \,
   {2 \over 3}
   \left ( {E_p + M_B \over 2 E_p} \right ) e^{- E_p t} .
   \label{gbb33}
\end{eqnarray}
\end{mathletters}
In determining the appropriate forms suitable for calculations using
Sakurai's conventions the definitions of the $\gamma$-matrices used in
the interpolating fields are taken into account.  Since the
nonvanishing terms of $G^{BB}_{\sigma \tau}$ are diagonal in $\sigma$
and $\tau$, the $\gamma$-matrices are paired with their Hermitian
conjugates.  Since the $\gamma$-matrix notations differ only by
factors of $i$ and $-1$, there are no required alterations for
calculations using the notation of Sakurai.

   Equations (\ref{gbb00}) through (\ref{gbb33}) provide four
correlation functions from which a baryon mass may be extracted.  All
baryon masses extracted from the different selections of Lorentz
indices agree within statistical uncertainties.  The combination
providing the smallest statistical fluctuations is $\bigm <
G^{BB}_{22} (t;\vec{p}, \Gamma_4) + G^{BB}_{33} (t;\vec{p},
\Gamma_4) \bigm >$ and these results are presented in section III.

   It should be noted that the spin-3/2 interpolating field also has
overlap with spin-1/2 baryons.  For the $\Delta$ baryons and
$\Omega^-$ this poses no problem as these baryons are the lowest lying
baryons in the mass spectrum having the appropriate isospin and
strangeness quantum numbers.  However, $\Sigma^*$ and
$\Xi^*$ correlation functions may have lower lying spin-1/2
components and therefore it is desirable to use the spin-3/2
projection operator \cite{benmerrouche89}
\FL
\begin{equation}
P^{3/2}_{\mu \nu}(p) = g_{\mu \nu} -
                      {1 \over 3} \gamma_\mu \gamma_\nu -
{1 \over 3 p^2} \left ( \gamma \cdot p \, \gamma_\mu \, p_\nu + p_\mu
\, \gamma_\nu \, \gamma \cdot p \right ) .
\label{proj}
\end{equation}
However, to use this operator, one must calculate the full $4^4$
matrix in Dirac and Lorentz spaces of $G^{BB}_{\sigma \tau}
(t;\vec{p}, \Gamma)$ which exceeds our current analysis of 4 Lorentz
terms and 2 Dirac terms by a factor of 32.

   QCD sum rule investigations of $\Sigma^*$ and $\Xi^*$ hyperons
suggest that the spin-1/2 component of the spin-3/2 interpolating
field is small relative to the spin-3/2 component \cite{belyaev83}.
However, the analysis does not determine whether the spin-1/2
component lies above or below the lowest lying spin-3/2 state.  Our
lattice results for baryon two-point functions give no indication of a
low-lying spin-1/2 component being excited by the spin-3/2
interpolating fields.

\subsection{Three-point functions and multipole form factors}

   Here we begin with a brief overview of the results of Ref.\
\cite{nozawa90}, where the multipole form factors are defined in
terms of the covariant vertex functions and in terms of the current
matrix elements.  The Dirac representation of the $\gamma$-matrices as
defined in Itzykson and Zuber continues to be used to facilitate
calculations of the $\gamma$-matrix algebra.  Finally, the results are
reported in Sakurai's notation in a form suitable for calculation in
lattice field theory.

   The electromagnetic current matrix element for spin-3/2 particles
may be written as
\widetext
\begin{equation}
\bigm < p',s' \bigm | j^{\mu}(0) \bigm | p,s \bigm > =
\sqrt{ M_B^2 \over E_p E_{p'} } \,
\overline{u}_{\alpha}(p',s') {\cal O}^{\alpha \mu \beta}
u_{\beta}(p,s) .
\label{cvf}
\end{equation}
Here $p,\ p'$ denote momenta, $s\ s'$ spins, and $u_\alpha(p,s)$ is a
Rarita-Schwinger spin-vector.  The following Lorentz covariant form
for the tensor
\begin{equation}
 {\cal O}^{\alpha \mu \beta} =
- g^{\alpha \beta} \left \{ a_{1} \gamma^{\mu}
+ {a_{2} \over 2 M_B} P^{\mu} \right \}
- \frac{ q^{\alpha} q^{\beta} }{(2M_{B})^{2}}
 \left\{ c_{1} \gamma^{\mu} + \frac{c_{2}}{2M_{B}} P^{\mu} \right\} ,
\label{tensor}
\end{equation}
where $P = p'+p$, $q=p'-p$ and $M_{B}$ is the mass of the baryon,
satisfies the standard requirements of invariance under time reversal,
parity, G-parity and gauge transformations.  The parameters $a_{1}$,
$a_{2}$, $c_{1}$ and $c_{2}$ are independent covariant vertex function
coefficients which are related to the multipole form factors.

  The multipole expansion of the electromagnetic current matrix
element, defined in terms of angular momentum recoupling algebra, has
the following form
\begin{mathletters}
\begin{eqnarray}
\lefteqn{\bigm < p',s' \bigm | j^{0}(0) \bigm | p,s \bigm > =}
\qquad \nonumber \\
&&A \bigm < \frac{3}{2} \, s' \bigm | G_{E0}(q^{2})
  + 2 \sqrt{5} \tau G_{E2}(q^{2}) \,
  \left[ \Sigma^{(2)}
  \times [\widehat{q} \times \widehat{q}]^{(2)} \right]^{(0)}
  \bigm | \frac{3}{2} \, s \bigm > \, ,  \label{mege}
\end{eqnarray}
\begin{eqnarray}
\lefteqn{\bigm < p',s' \bigm | \vec{j}(0) \bigm | p,s \bigm > =}
\qquad \nonumber \\
&&\sqrt{\tau}
    \bigm < \frac{3}{2} \, s' \bigm |  \left\{ G_{E0}(q^{2})
  + 2 \sqrt{5} \tau G_{E2}(q^{2}) \,
    \left[ \Sigma^{(2)} \times [\widehat{q}
    \times \widehat{q}]^{(2)} \right]^{(0)}
    \right\}  \widehat{P}  \nonumber \\
&&+ \, i \left\{ \, \frac{1}{3} G_{M1}(q^{2}) \Sigma^{(1)}  \,
  + \, 3 \tau G_{M3}(q^{2}) \left[ \Sigma^{(3)} \times
  \left[ \widehat{q} \times \widehat{q}
  \right]^{(2)} \right]^{(1)} \, \right\}
  \times \widehat{q} \bigm | \frac{3}{2} \, s \bigm > \, ,
  \label{megm}
\end{eqnarray}
\end{mathletters}
where $\tau = - q^{2}/(2M_{B})^{2}$ $(\geq 0)$, and $\widehat{P}$ and
$\widehat{q}$ are unit vectors.  $A = \sqrt{1 + \tau}$ in the
laboratory frame ($\vec{p}=0$) and $A= 1$ in the baryon Breit frame
($\vec{P}= \vec{p} \,' + \vec{p} = 0$).  The spin matrix elements are
defined by Clebsch-Gordan coefficients,
\begin{mathletters}
\begin{eqnarray}
&&\bigm < \frac{3}{2} \, s' \bigm | \frac{3}{2} \, s \bigm > =
  \delta_{s's} \, , \\
&&\bigm < \frac{3}{2} \, s' \bigm |
  \Sigma^{(1)}_{m} \bigm | \frac{3}{2} \, s \bigm >
= \sqrt{15} \, \left ( \frac{3}{2} s' 1 m
  \biggm | \frac{3}{2} 1 \frac{3}{2} s \right ) \, , \\
&&\bigm < \frac{3}{2} \, s' \bigm |
  \Sigma^{(2)}_{m} \bigm | \frac{3}{2} \, s \bigm >
= - \sqrt{\frac{5}{6}} \, \left ( \frac{3}{2} s' 2 m
    \biggm | \frac{3}{2} 2 \frac{3}{2} s \right ) \, , \\
&&\bigm < \frac{3}{2} \, s' \bigm |
  \Sigma^{(3)}_{m} \bigm | \frac{3}{2} \, s \bigm >
= - \frac{7}{6} \sqrt{\frac{2}{3}} \,
    \left ( \frac{3}{2} s' 3 m \biggm |
    \frac{3}{2} 3 \frac{3}{2} s \right ) \, ,
\end{eqnarray}
\end{mathletters}
where the Condon and Shortley phase convention has been used
\cite{edmonds60}.

   The multipole form factors are defined in terms of the covariant
vertex function coefficients $a_{1}$, $a_{2}$, $c_{1}$ and $c_{2}$
through the following Lorentz invariant expressions \cite{nozawa90},
\begin{mathletters}
\begin{eqnarray}
{\cal G}_{E0}(q^{2}) & = & ( 1 + \frac{2}{3} \tau ) \left\{
                       a_{1} + (1+\tau) a_{2} \right\}
 - \frac{1}{3} \tau (1+\tau) \left\{ c_{1} +
                    (1+\tau)c_{2} \right\} \, , \\
{\cal G}_{E2}(q^{2}) & = & \left\{ a_{1} + (1+\tau) a_{2} \right\}
 - \frac{1}{2} (1+\tau) \left\{ c_{1} +
               (1+\tau)c_{2}  \right\} \, , \\
{\cal G}_{M1}(q^{2}) & = & ( 1 + \frac{4}{5} \tau ) a_{1}
                       - \frac{2}{5} \tau (1+\tau) c_{1} \, , \\
{\cal G}_{M3}(q^{2}) & = &  a_{1} - \frac{1}{2} (1+\tau) c_{1} \, .
\end{eqnarray}
\end{mathletters}
The multipole form factors ${\cal G}_{E0}$, ${\cal G}_{E2}$, ${\cal
G}_{M1}$ and ${\cal G}_{M3}$ are referred to as charge ($E0$),
electric-quadrupole ($E2$), magnetic-dipole ($M1$) and
magnetic-octupole ($M3$) multipole form factors, respectively.

   In a manner similar to that for the two-point function, the
three-point Green function for the electromagnetic current is defined
as
\begin{eqnarray}
\lefteqn{
\bigm < G^{B j^\mu B}_{\sigma \tau}
               (t_2, t_1; \vec{p'}, \vec{p}; \Gamma) \bigm > \,
   = } \qquad
\nonumber \\
& & \sum_{\vec{x_2}, \vec{x_1}}
   e^{-i \vec{p'} \cdot \vec{x_2} } e^{+ i \left (
   \vec{p'} - \vec{p} \right ) \cdot \vec{x_1} }
   \Gamma^{\beta \alpha} \bigm < \Omega \bigm | T
   \left ( \chi^\alpha_\sigma(x_2) j^\mu(x_1)
   \overline \chi^\beta_\tau(0) \right )
   \bigm | \Omega \bigm > .
\end{eqnarray}
Once again, the subscripts $\sigma, \, \tau$ are the Lorentz indices
of the spin-3/2 interpolating fields.  For large Euclidean time
separations $t_2 - t_1 >\!> 1$ and $t_1 >\!> 1$ the three-point
function at the hadronic level takes the limit
\begin{eqnarray}
\lefteqn{
\bigm < G^{B j^\mu B}_{\sigma \tau} (t_2, t_1;\vec{p'}, \vec{p};
                                     \Gamma) \bigm > =
\sum_{s, s^\prime} e^{-E_{p'} (t_2-t_1)} e^{-E_p t_1} }
\qquad \nonumber  \\
& & \Gamma^{\beta \alpha} \bigm < \Omega \bigm |
    \chi^\alpha_\sigma \bigm | p', s^\prime \bigm >
    \bigm < p', s^\prime \bigm |
    j^\mu \bigm | p, s \bigm > \bigm < p, s \bigm |
    \overline \chi^\beta_\tau  \bigm | \Omega \bigm > .
\end{eqnarray}
where the matrix element of the electromagnetic current is defined in
(\ref{cvf}), and the matrix elements of the interpolating fields are
defined by (\ref{interp}).

   In Ref.\ \cite{nozawa90} it was noted that the time dependence of
the three-point function may be eliminated by use of the two-point
functions.  However the appropriate combination of the two-point
function Lorentz indices was not specified.  Maintaining the lattice
Ward identity, which guarantees the lattice electric form factor
reproduces the total charge of the baryon at $q^2=0$, provides an
indispensable guide to the optimum ratio of Green functions.  The
preferred ratio of two- and three-point Green functions is
\begin{eqnarray}
\lefteqn{
   R_\sigma{}^\mu{}_\tau (t_2,t_1; \vec{p'}, \vec{p};
                          \Gamma ) = } \qquad \nonumber \\
   & & \left ( {
\bigm < G^{B j^\mu B}_{\sigma \tau}
      (t_2,t_1; \vec{p'}, \vec{p}; \Gamma) \bigm >
\bigm < G^{B j^\mu B}_{\sigma \tau}
      (t_2,t_1; -\vec{p}, -\vec{p'}; \Gamma) \bigm >
\over
\bigm < G^{BB}_{\sigma \tau} (t_2; \vec{p'}; \Gamma_4 ) \bigm >
\bigm < G^{BB}_{\sigma \tau} (t_2; -\vec{p}; \Gamma_4 ) \bigm > }
\right )^{1/2} , \label{ratio} \\
& & \simeq
   \left ( { E_p + M \over 2 E_p } \right )^{1/2}
   \left ( { E_{p'} + M \over 2 E_{p'} } \right )^{1/2}
   \overline R_\sigma{}^\mu{}_\tau (\vec{p'}, \vec{p}; \Gamma )
   \, , \label{redratio}
\end{eqnarray}
where we have defined the reduced ratio $\overline
R_\sigma{}^\mu{}_\tau (\vec{p'}, \vec{p}; \Gamma )$.  Note that there
is no implied sum over $\sigma$ and $\tau$ in (\ref{ratio}).

   Using our standard definitions for $\Gamma$ given in
(\ref{gammadef}) and the Rarita-Schwinger spin sum of (\ref{rsss}),
the multipole form factors may be isolated and extracted.  The
appropriate combinations of $\overline R_\sigma{}^\mu{}_\tau
(\vec{p'}, \vec{p}; \Gamma )$ suitable for calculations employing the
$\gamma$-matrix and metric conventions of Sakurai are
\begin{mathletters}
\begin{eqnarray}
{\cal G}_{E0}(q^2) &=& {1 \over 3} \left (
                      \overline R_1{}^4{}_1(\vec q_1,0; \Gamma_4)
                 +    \overline R_2{}^4{}_2(\vec q_1,0; \Gamma_4)
                 +    \overline R_3{}^4{}_3(\vec q_1,0; \Gamma_4)
                      \right ) , \label{ge0} \\
{\cal G}_{E2}(q^2) &=& 2{M (E+M) \over |\vec q_1|^2} \left (
                      \overline R_1{}^4{}_1(\vec q_1,0; \Gamma_4)
                 +    \overline R_2{}^4{}_2(\vec q_1,0; \Gamma_4)
               - 2 \, \overline R_3{}^4{}_3(\vec q_1,0; \Gamma_4)
                      \right ) , \nonumber \\
&& \label{ge2} \\
{\cal G}_{M1}(q^2) &=& - \, {3 \over 5}{E+M \over |\vec q_1|}
                       \left (
                      \overline R_1{}^3{}_1(\vec q_1,0; \Gamma_2)
                 +    \overline R_2{}^3{}_2(\vec q_1,0; \Gamma_2)
                 +    \overline R_3{}^3{}_3(\vec q_1,0; \Gamma_2)
                      \right ) , \label{gm1} \\
{\cal G}_{M3}(q^2) &=& - \, 4 {M (E+M)^2 \over |\vec q_1|^3}
                      \left (
                      \overline R_1{}^3{}_1(\vec q_1,0; \Gamma_2)
                 +    \overline R_2{}^3{}_2(\vec q_1,0; \Gamma_2)
   -    {3 \over 2}\, \overline R_3{}^3{}_3(\vec q_1,0; \Gamma_2)
                      \right ) , \nonumber \\
&& \label{gm3}
\end{eqnarray}
\end{mathletters}
\narrowtext
\noindent
where $\vec q_1 = (q,0,0)$.  Equation (\ref{ge2}) for ${\cal G}_{E2}$
isolates the spin matrix element \break $\bigm < {3 \over 2} s'
\bigm | \Sigma^{(2)}_0 \bigm | {3 \over 2} s \bigm >$.  Smaller
statistical uncertainties may be obtained by using the symmetry
$\overline R_2{}^4{}_2(\vec q_1,0; \Gamma_4) = \overline
R_3{}^4{}_3(\vec q_1,0; \Gamma_4)$.  This symmetry is used in the
following analysis to eliminate the term $\overline R_2{}^4{}_2(\vec
q_1,0; \Gamma_4)$ in the $E2$ form factor.  The signs of the $E2$ form
factors remain unchanged.

   It is worth noting at this point that the form of the tensor ${\cal
O}^{\alpha \mu \beta}$ in (\ref{tensor}) is not unique.  While
equations (\ref{ge0}) through (\ref{gm3}) are derived using
(\ref{tensor}) these results are more generally applicable.  For
example, the authors of Ref.\ \cite{korner77} employ the form
\begin{eqnarray}
 {\cal O}^{\alpha \mu \beta} &=& g^{\alpha \beta}
\left \{ F_{1} \gamma^{\mu}
+ {F_{2} \over 2 M_B} i \sigma^{\mu \nu} q_{\nu} \right \}
\nonumber \\
&& + \frac{ q^{\alpha} q^{\beta} }{2 M_{B}^{2}}
 \left\{ F_{3} \gamma^{\mu} +
 \frac{F_{4}}{2 M_{B}} i \sigma^{\mu \nu} q_{\nu} \right\} .
\end{eqnarray}
Provided one includes a factor of $\sqrt{6}$ to account for their
normalization of the $M3$ form factor, the results of (\ref{ge0})
through (\ref{gm3}) isolate the multipole form factors.

\subsection{Lattice techniques}

   Here we briefly summarize the lattice techniques used in the
following calculations.  Additional details may be found in Ref.\
\cite{leinweber91}.

   We use Wilson's formulation for both the gauge and fermionic
action.  $SU(2)$-isospin symmetry is enforced by equating the Wilson
hopping parameters $\kappa_u = \kappa_d = \kappa$.  We select three
values of $\kappa$, which we denote $\kappa_1 = 0.152$, $\kappa_2 =
0.154$ and $\kappa_3 = 0.156$, and extrapolate the $u$-$d$ quark
sector to the chiral limit.  To account for the relatively heavy
strange quark we fix $\kappa_s = \kappa_1$, the smallest of the three
values of $\kappa$ considered.  This allows an acceptable
extrapolation of the light quarks to the chiral limit through values
of quark mass less than or equal to the strange quark mass. Our
calculations of octet and decuplet baryon masses indicate that this
selection for $\kappa_s$ gives a reasonable description of the strange
quark dynamics.

   The conserved electromagnetic current is derived from the fermionic
action by the Noether procedure.  The lattice Ward identity guarantees
the lattice electric form factor reproduces the total baryon charge at
$q^2=0$.  The quark propagators coupled with fixed momentum $\vec q_1
= (q,0,0)$ to $j^\mu$ are calculated using the sequential source
technique (SST) \cite{bernard86,bernard84,draper84}.

   To minimize noise in the Green functions, we exploit the parity
symmetry of the correlation functions, and the equal weighting of
$\{U\}$ and $\{U^*\}$ gauge configurations in the lattice action.
Defining $s_P$ as
\FL
\begin{equation}
G(\vec{p'},\vec p, \vec q; \Gamma)
     = s_P \, G(-\vec{p'},-\vec p,-\vec q; \Gamma), \qquad s_P=\pm 1 ,
\end{equation}
and $s_C$ as
\begin{equation}
\Gamma = s_C \left ( \widetilde C \Gamma \widetilde C^{-1} \right )^*
, \qquad s_C=\pm 1 ,
\end{equation}
where $\widetilde C = C \gamma_5$, the correlation functions are real
provided
\begin{equation}
s_C = s_P.
\end{equation}
This condition is satisfied with the selections for $\Gamma$ indicated
in (\ref{ge0}), though (\ref{gm3}).  While this approach requires an
extra matrix inversion to determine an additional SST propagator with
momentum $-\vec q_1$, inclusion of both $\{U\}$ and $\{U^*\}$
configurations in the calculation of the correlation functions
provides an unbiased estimate of the ensemble average properties which
has substantially reduced fluctuations
\cite{draper89p}.

Twenty-eight quenched gauge configurations are generated by the
Cabibbo-Marinari \cite{cabibbo82} pseudo-heat-bath method on a $24
\times 12 \times 12 \times 24$ periodic lattice at $\beta=5.9$.
Dirichlet boundary conditions are used for fermions in the time
direction.  Configurations are selected after 5000 thermalization
sweeps from a cold start, and every 1000 sweeps thereafter
\cite{correlations}.  Time slices are labeled from 1 to 24, with the
$\delta$-function source at $t=4$. A symmetric combination of the
current $( j^\mu(x_1-\widehat \mu) + j^\mu(x_1) )/2$ is centered at
time slice $t_1=12$. The following calculations are done in the lab
frame $\vec p=0,\; \vec{p'}= \vec q_1 = (2\pi/24,0,0)$, the minimum
nonzero momentum available on our lattice. The spatial direction of
the electromagnetic current is chosen in the $z$-direction.  As
discussed in section III B and as in Ref.\ \cite{leinweber91}, the
reported masses and form factors are determined by fitting the
correlation functions in time slices 16 through 20 inclusive.

Statistical uncertainties are calculated in a third-order, single
elimination jackknife \cite{efron79,gottlieb86}.  A third order
jackknife provides uncertainty estimates for the correlation
functions, fits to the correlation functions, and quantities
extrapolated to the chiral limit.

\section{RESULTS}

\subsection{Baryon mass}

   Baryon masses are determined by fitting the Euclidean time
evolution of baryon two-point functions at three values of $\kappa$
with the strange quark fixed at the smallest value of $\kappa$.  The
light $u$ and $d$ quarks are extrapolated to $\kappa_{cr}$ where an
extrapolation of the squared pion mass vanishes.  The nucleon mass is
used to define the lattice spacing $a=0.128(11)$ fm, $a^{-1} =
1.54(13)$ GeV.  Figure \ref{decmass} displays the masses of the
decuplet baryons obtained from the combination of two-point functions
$G^{BB}_{22} + G^{BB}_{33}$.  Octet baryon masses are also displayed
for reference.  Experimental masses \cite{pdt} are indicated by
horizontal lines.  With the exception of the $N$-$\Delta$ splitting,
the baryon masses are reasonably reproduced.  Table \ref{baryonmass}
summarizes the lattice baryon masses at the three values of $\kappa$
considered as well as the extrapolated masses.  The momentum transfer
is relatively insensitive to the baryon mass.  For all baryons, $q^2
a^2 = 0.067 \pm 0.001$ with the larger values corresponding to the
heavier baryons.

\subsection{Correlation function ratios}
\label{corrfr}

   Here we present a few of the correlation function ratios of
(\ref{ratio}) which are combined to isolate the multipole form
factors.  Figure \ref{sige0R3} displays the quark contributions to the
electric charge form factor of $\Sigma^*$ hyperons at $\kappa_u =
\kappa_3$, the lightest quark masses considered.  The $d$-quark
contribution is related to the $u$-quark result by a ratio of the
quark charges.  Electric charge form factors for the three charge
states of the $\Sigma^*$ hyperons are obtained through the appropriate
sums of quark contributions.  Similar results are seen for other
decuplet baryons.

   The form factors are determined by fitting the correlation function
ratio sum by a horizontal line for times $t_2 >> 12$ and $t_2 << 24$.
Fortunately there is a rather broad plateau region where the electric
form factor may be reliably determined.  We consider fits of the
correlation functions from time slice 15 through 21 in intervals
including 4 to 7 points.  The results are selected from these 10 fits
based on the flatness of the correlation functions and the statistical
uncertainties. As in the octet baryon analysis, it is found that fits
of the 5 points in the time slice interval 16 to 20 provide the
optimum balance between these systematic and statistical
uncertainties.

   Figure \ref{xim1R3} shows a similar plot for the quark
contributions to the magnetic-dipole form factor of $\Xi^*$ hyperons.
Once again $\kappa_u = \kappa_3$ with $\kappa_s$ fixed at
$\kappa_1$.

   Figures \ref{dele2e0R3} and \ref{delm3m1R3} display the $E2$ and
$M3$ ratio sums for $\Delta$ baryons in comparison with $E0$ and $M1$
ratio sums respectively.  Only the ratio terms appearing inside the
parentheses of (\ref{ge0}) through (\ref{gm3}) are included here.  The
higher-order moments are dependent upon subtle differences in the
ratio contributions, $R_\sigma{}^\mu{}_\tau$.  Fortunately the
statistical uncertainties of individual correlation function ratios
are correlated and some of the uncertainty is canceled in the
higher-order moment sums.  However, statistical uncertainties in the
$E2$ and $M3$ moments remain large.

   The $E2$ result is small relative to the $E0$ ratio sum.  In the
interest of placing an upper limit on the magnitude of the $E2$ form
factor we will consider fits of the $E2$ correlation functions in the
standard interval of $t_2 = 16$ through 20.  The $M3$ moment is finite
and remains finite for all three values of $\kappa$ considered.  This
is not entirely surprising as pure $M1$ multipole moments are
generally not seen in extended physical systems.  Unfortunately, a
clear plateau region is not seen.  To determine the scale of the
$M3$ moment we will fit the correlation function ratio sum in the
standard interval of $t_2 = 16$ through 20 with the understanding that
the result may be an underestimate of the $M3$ moment.

\subsection{Electric charge form factors}

   Electric charge form factors are extracted from fits of the ratios
in (\ref{ge0}).  Table \ref{baryone0ff} reports the form factors for
the decuplet baryons at the three values of $\kappa$ considered along
with the extrapolated values.  Table \ref{quarke0ff} gives the form
factors for individual quarks of unit charge within decuplet baryons.
Note that isospin symmetry equates $d$-quark properties to $u$-quark
properties when normalized to unit charge.

   We use the standard small $q^2$ expansion of the Fourier transform
of a spherical charge distribution to extract the electric charge
radius.
\begin{equation}
\bigm < r^2 \bigm > =
-6 {d \over dq^2} {\cal G}_E(q^2) \bigm |_{q^2=0} .
\label{rsq}
\end{equation}
We have two points describing the function ${\cal G}_E(q^2)$. Namely
${\cal G}_E(0)$, the total charge of the baryon, and ${\cal G}_E(q^2)$
evaluated at the smallest finite value of $q^2$ available on our
lattice. To evaluate the derivative of (\ref{rsq}) we use a dipole
form for the function ${\cal G}_E(q^2)$.  The dipole result for the
radius is
\begin{equation}
{\bigm < r^2 \bigm > \over {\cal G}_E(0)} = {12 \over q^2} \left (
    \sqrt{ {\cal G}_E(0) \over {\cal G}_E(q^2) } - 1 \right ).
\label{dipole}
\end{equation}
To assess the sensitivity of our results on the dipole approximation
we also consider a monopole form.

   In the figures and tables of this section we quote the quantity
$\sqrt{ \bigm < r^2 \bigm > / {\cal G}_E(0) }$ which gives the radius
of baryons and quark distributions with unit charge.  In all cases the
sign of $\bigm < r^2 \bigm >$ is the same as the charge of the baryon
or quark.  The difference of the radii extracted in the dipole and
monopole approximations is small relative to the statistical
uncertainties in the radii.  We refer to the dipole results in the
following discussion and figures.

   The charge radii of decuplet baryons are not determined by
experiment so our goal here is to compare decuplet and octet results
to see if there is some indication of how spin-dependent interactions
affect the electromagnetic structure.  Of course the spin-dependent
interactions depend on the quark masses and our calculations are done
at masses which are not yet small.  However the extrapolation of
charge radii into the physical (light) quark region is problematic
since charge radii become infinite in the exact chiral limit
\cite{beg72}.  To reproduce this behavior in the present calculation
would require some model dependent theoretical input which is not
available.

   To estimate the charge radii in the physical mass region and allow
a comparison with model calculations which often neglect couplings to
the continuum, we extrapolate in $1/\kappa$ (or equivalently in
$m_\pi^2$) to $\kappa_{cr}$.  Although this prescription involves some
systematic uncertainty \cite{leinweber92c}, it gives a better picture
of what is happening at physical quark masses than, for example, using
only results calculated at our largest $\kappa$ value.  However it
should be noted that the qualitative statements and conclusions are
the same whether we extrapolate or not.

   Figure \ref{chgrad} displays the lattice results for the
electric charge radii normalized to the proton charge radius for the
charged members of the baryon octet and decuplet.  We find a similar
pattern of quark mass effects in the octet and decuplet with a
$\Delta^+$ charge radius essentially equal to that of the proton.
Note that the charge radii of $\Delta^{++}$ and $\Delta^-$ are equal
to that of $\Delta^+$ when the radii are normalized by the total
charge.  Table \ref{baryonchgrad} summarizes the lattice calculations.

   Some insight into quark dynamics may be obtained by examining the
behavior of the quark distribution radii in hyperons as the $u$ and
$d$ quarks become lighter.  Consider for example, the distribution
radii of $u$ and $s$ quarks within $\Sigma^+$ as the $u$ quarks are
extrapolated to $\kappa_{cr}$.  Figures \ref{octsigqr} and
\ref{decsigqr} display the quark distribution radii in lattice units
(LU) for the octet $\Sigma^+$ and the decuplet $\Sigma^{*+}$
respectively.

   In both cases, the radius of the $u$-quark distribution increases
as the $u$-quark becomes lighter (smaller $\kappa^{-1}$), as expected.
However there is a difference in the behavior of the $s$-quark
distributions in the octet $\Sigma^+$ and the decuplet $\Sigma^{*+}$.
In the case of octet $\Sigma^+$, there is a small, but significant,
trend for the $s$-quark distribution radius to decrease as the
$u$-quarks become lighter \cite{raddiff}.  This effect was understood
as a shifting of the center of mass of the $u$-$s$ system towards the
strange quark which becomes relatively heavier as the $u$-quarks
become lighter.  In contrast, the $s$-quark distribution radius in the
decuplet $\Sigma^{*+}$ has little dependence on the $u$-quark mass.

   This difference between octet and decuplet behavior has a natural
explanation in the hyperfine interaction term of the
one-gluon-exchange potential.  The hyperfine force is repulsive for
quarks interacting in spin-triplet states.  The strength of the
interaction increases with decreasing $u$-quark mass and provides a
mechanism to counteract the center-of-mass effect.

   Lattice results for the neutron electric form factor confirm a
repulsive force between quarks with their spins aligned.  Furthermore,
the magnetic form factors indicate that in the octet $\Sigma^+$
(decuplet $\Sigma^{*+}$), the singly represented $s$ quark has, on
average, its spin anti-aligned (aligned) with that of the doubly
represented $u$ quarks.  Hence the spin alignments of $u$ and $s$
quarks seen in the lattice results are in qualitative agreement with
those anticipated from $SU(6)$-spin-flavor symmetry.  Note however,
the lattice dynamics do not impose this symmetry.  The quark
contributions to octet baryon magnetic moments differ significantly
{}from the $SU(6)$ predictions.

    Figure \ref{qrkradii} shows charge distribution radii for quarks
of unit charge in decuplet baryons.  The analogous graph for octet
baryons is indicated in figure \ref{oqchgrad}.  It should be noted
that although the uncertainty regions of the radii for different
quarks overlap it does not necessarily mean that the $u$ quark
distribution radius in $n$, for example, may be larger than the $u$
quark radius in $\Xi^0$.  The uncertainties are highly correlated
between the two results and a calculation of the difference of the
radii indicates $u_\Xi$ is larger by $0.50{+0.65 \atop -0.15}$~LU.
Other octet baryon quark pairs such as $u_\Sigma - u_p$ differ from
zero by at least one standard deviation.  In the octet, therefore,
some nontrivial baryon dependence of the quark distributions does
occur.  In contrast, significant baryon dependence of the quark
distributions in decuplet baryons is not observed.  The lattice
results are summarized in table \ref{quarkchgrad}.

   Figure \ref{qchgradrat} displays ratios of octet and decuplet quark
charge distribution radii.  With the exception of the singly
represented octet quarks $u_n$ and $s_\Sigma$ which showed some baryon
dependence in their charge distributions, the remaining quark
distribution radii are unaffected by differences between octet and
decuplet spin-flavor symmetry.

   Figure \ref{qchgradrat} also gives an understanding as to why the
proton and $\Delta^+$ have similar charge radii.  Naively one might
expect $\Delta^+$ to be larger due to additional spin-dependent
repulsion between the quarks.  However, the dominant effect of the
additional spin repulsion is to enhance the charge distribution radius
of the $d$ quark in $\Delta^+$ relative to that in $p$.  This
counteracts any possible enhancement of $u$-quark radii in $\Delta^+$.

   Some attention has been given to model calculations of the
$\Omega^-$ charge radius.  The model predictions vary over a wide
range.  Figure \ref{omegarad} illustrates results for the $\Omega^-/p$
charge radius ratio.  The results include the lattice (Latt.) results
of this analysis, a calculation based on the relativised quark model
(Q.M.) of Ref.\ \cite{barik90}, MIT bag (MIT) and cloudy bag (C.B.)
models \cite{kunz91} and a bound state approach Skyrme (Skyr.) model
calculation \cite{kunz90}.  Only the relativised quark model
calculation agrees with the lattice results.

   The electric form factors calculated at $q \simeq 0.4$ GeV for
neutral decuplet baryons are illustrated in figure \ref{e0neut}.
Neutral octet baryons are also given for reference.  For the hyperons
the form factors are dominated by the net charge of the light quarks.
Octet and decuplet hyperons display similar behavior.  For $\Delta^0$
the symmetric spin state of the quarks causes the form factor to
vanish.  This contrasts the neutron where the octet-spin asymmetry of
the three quarks gives rise to a negative squared charge radius.

   With knowledge of quark distribution radii, the charge radii of
neutral baryons may be calculated as in $r^2 = \sum_{i=s,s,u} e_i \,
r_i^2$ for $\Xi^0$.  Table \ref{neutchgrad} summarizes the results.
The lattice result for the squared charge radius of the neutron
$r_n^2/r_p^2 = -0.11{+0.10 \atop -0.14}$ encompasses the experimental
value \cite{hohler76,dumbrajs83} of $-0.167(7)$.

\subsection{Magnetic-dipole form factors}

   Our calculation of magnetic-dipole form factors is done at the
smallest finite value of $q^2$ available on our lattice.  Table
\ref{baryonm1ff} summarizes the form factors in units of natural
magnetons $(e/2 M_B)$ where the mass of the baryon appears in the
definition of the magneton.  The magnetic moment $\mu$ is defined at
$q^2 = 0$ as ${\cal G}_{M1}(0) = \mu / (e/2M_B)$ and therefore we must
scale our results from ${\cal G}_{M1}(q^2)$ to ${\cal G}_{M1}(0)$.
Lattice extrapolations in $q^2$ to $q^2 = 0$ suffer from large
statistical errors. To make contact with the experimental magnetic
moments, we assume equal scaling of electric and magnetic form factors
as in the octet baryon analysis and as supported by experimental
measurements of nucleon form factors.  In hyperons, the strange and
light quark sectors are scaled separately by
\begin{equation}
{ {\cal G}_{M1}^s(0) \over {\cal G}_{M1}^s(q^2) } =
{ {\cal G}_E^s   (0) \over {\cal G}_E^s   (q^2) } ,
\end{equation}
and similarly for the light quarks, such that the magnetic moment of
a hyperon is given by
\begin{equation}
{\cal G}_{M1}^B(0) = {\cal G}_{M1}^l(0) + {\cal G}_{M1}^s(0),
\end{equation}
where $l$ labels the light quarks. For $\Delta$ baryons it is not
necessary to separate the $u$- and $d$-quark sectors due to the
$SU(2)$-isospin symmetry of the correlation functions.  Table
\ref{quarkm1ff} gives the magnetic-dipole form factors for quarks of
unit charge in units of natural magnetons.  Combined with the results
of table \ref{quarke0ff} the magnetic moments may be calculated.

   In the octet baryon analysis it was found that the magnitudes of
the lattice results for magnetic moments were consistently smaller
than the experimental measurements.  It was argued that at $\beta=5.9$
some deviations from asymptotic scaling may occur.  A recent analysis
\cite{wilcox92} determines nucleon form factors at $\beta=6.0$ on a
cubic lattice with physical spatial dimensions roughly equal to our
smaller $y$ and $z$ dimensions.  Some improvement is seen in the
magnitudes of the magnetic moments which are still 10 to 15\% low
compared to experiment.  Chiral dressings of the nucleon may cause our
linear extrapolation of the magnetic moments in $1/\kappa$ to
underestimate the magnetic moments in the physical regime
\cite{leinweber92c}.  Finite volume effects may also give rise to the
underestimation of the magnetic moments as the baryon is restricted by
its periodic images.  The proton rms electric charge radius at
$\kappa_3$ indicates the proton largely fills the lattice in our
smaller $y$ and $z$ spatial dimensions.  Non-quenched corrections may
also provide additional contributions \cite{leinweber92a}.

   To reduce the effects of these uncertainties, ratios of the lattice
results to the lattice proton result are used when making comparisons
with experimental measurements or model calculations.  Table
\ref{baryonmagmom} reports the extrapolated decuplet baryon magnetic
moments in units of nuclear magnetons $(\mu_N)$, ratios of decuplet
baryon moments and the proton magnetic moment and ratios scaled to
reproduce the proton magnetic moment.  These scaled ratios are
illustrated in figure \ref{decmagmqm}.  The expected qualitative
behavior of mass dependence is displayed here.  For example as $d$
quarks are replaced by $s$ quarks in going from $\Delta^-$ to
$\Sigma^{*-}$ through to $\Omega^-$ the magnetic moments of the
negatively charged hyperons decrease in magnitude.  As in the case of
the electric charge form factors of neutral baryons, the magnetic
moments are dominated by the net charges of the light quarks.
$SU(2)$-isospin symmetry causes the $\Delta^0$ moment to vanish.

   The simple quark model formula for the magnetic moment of a
decuplet baryon is simply given by the sum of the intrinsic moments of
the quarks composing the baryon.  Results using intrinsic moments
determined by the $p$, $n$, and $\Lambda$ moments \cite{pdt} are also
indicated in figure \ref{decmagmqm} by horizontal dashed lines.  The
agreement is striking and suggests a baryon independence of the
lattice effective quark moments.

   Figure \ref{qrkmagm} displays the lattice effective quark moments
for quarks of unit charge within decuplet baryons.  Effective moments
have been defined by equating the lattice quark sector contributions
to the same sector of the $SU(6)$-magnetic-moment formula derived from
$SU(6)$-spin-flavor symmetry wave-functions.  For example,
$SU(6)$-spin-flavor symmetry gives the simple quark model formula
\begin{equation}
\mu_p = A \mu_u - B \mu_d ,
\label{mup}
\end{equation}
for the magnetic moment of the proton, where $A=4/3$ and $B=1/3$.  The
effective moment of the lattice $u$-quark in the proton is defined by
equating the lattice $u$-quark sector contribution to ${4\over 3}
\mu_u$, the corresponding $u$-quark sector contribution of the simple
quark model. Similarly, the lattice $d$-quark sector contribution is
equated with $-{1\over 3} \mu_d$ in defining the effective $d$-quark
moment.  In $\Delta^+$ the lattice $u$-quark sector is equated with $2
\mu_u$ and the $d$-quark sector equals the effective $d$-quark moment.
Numerical values for effective quark moments are summarized in table
\ref{quarkeffmom} for both octet and decuplet baryons.

   Approximate baryon independence of decuplet baryon effective quark
moments is displayed in figure \ref{qrkmagm}.  Closer examination of
effective $u$-quark magnetic moment ratios does reveal some small
baryon dependence of the quark moments.  These results are summarized
in table \ref{eqmmr}.  The quenching of effective quark moments in
hyperons is largely due to the baryon mass setting the scale at which
quarks contribute to the baryon moment \cite{leinweber92a}.  These
results are in sharp contrast to the enormous baryon dependence of the
effective quark moments seen in octet baryons as illustrated in figure
\ref{octqmagmom}.

   It is instructive to consider ratios of the octet baryon effective
quark moments with their decuplet partners.  These ratios are
illustrated in figure \ref{qmagmomrat}.  In the simple quark model all
of these ratios are equal to 1.  Dramatic effects are seen for the
singly represented quarks $u_n$, $s_\Sigma$ and $u_\Xi$ where
deviations from $SU(6)$-octet wave-functions are large in the lattice
results.  All quarks show a significant baryon dependence with the
exception of $u_p/u_\Delta$ and possibly $u_\Sigma/u_{\Sigma^*}$.
This rich octet/decuplet baryon dependence in the effective quark
moments contrasts that of the quark charge distribution radii as
illustrated in the ratios of figure \ref{qchgradrat}.

   While lattice octet baryon moments differ significantly from the
predictions of $SU(6)$, it was not possible to determine how each of
the coefficients $A=4/3$ and $B=1/3$ of (\ref{mup}) are altered.
Using ratios of the $u$- and $d$-quark sector contributions to $p$ or
$n$ and isospin symmetry, the octet baryon analysis indicates
\begin{equation}
{B \over A} = 0.13{\textstyle{ +.05 \atop -.11}} < {1 \over 4} ,
\label{bona}
\end{equation}
where $1/4$ is the $SU(6)$ prediction.  The decuplet baryon results
indicate the effective moments and charge radii of doubly represented
quarks are largely unchanged suggesting $A \simeq 4/3$ and indicating
$B$ is better approximated by $1/6$ than the standard $SU(6)$ value of
$1/3$.

   Considerable effort has gone into model calculations of the
$\Delta^{++}$ and $\Omega^-$ magnetic moments.  Here we collect
together a handful of these calculations for comparison with our
lattice results.  Figure \ref{deltapp} displays ratios of
$\Delta^{++}/p$ magnetic moments.  The model calculations include
results from the simple quark model \cite{pdt} (Q.M.), cloudy bag
\cite{krivoruchenko87} (C.B.), Skyrme \cite{kim89} (Skyr.),
Bethe-Salpeter \cite{mitra84} (B.S.) and QCD sum rule
\cite{belyaev92} (S.R.)  analyses.  The experimental
result (Expt.) is from a recent pion bremsstrahlung analysis
\cite{bosshard91} and therefore has some model dependence.  Earlier
papers reported larger values \cite{nefkens78,wittman88} in better
agreement with our lattice result.

   Figure \ref{omegamin} displays ratios of $\Omega^-/p$ magnetic
moments.  The labels are as in figure \ref{deltapp} with additions of
an additive quark model \cite{chao90} in which effective masses (E.M.)
of the quarks are used to estimate the intrinsic quark moments and a
calculation in which relativistic corrections (R.C.) to baryon
magnetic moments are considered \cite{georgi83}.  Experimental results
(Expt.) are from a recent investigation where $\Omega^-$ hyperons are
produced by a polarized neutral beam spin transfer reaction
\cite{diehl91}.

   In the simple quark model the ratio of $\Omega^-/\Lambda^0$
magnetic moments is 3.  The lattice results suggest that
spin-dependent forces give rise to a larger strange quark moment in
$\Omega^-$ relative to that in $\Lambda^0$ as indicated in figure
\ref{qmagmomrat}.  The lattice ratio of $\Omega^-/\Lambda^0$ magnetic
moments is $3.6{+1.0 \atop -0.6}$ suggesting an enhancement over the
simple quark model ratio.

   To complete the discussion of decuplet baryon magnetic-dipole
properties figure \ref{dbmagrad} summarizes our calculation of
magnetic radii normalized by the magnetic moment as in $\sqrt{ < r^2 >
/ {\cal G}_{M1}(0)}$.  The magnetic radii follow a similar pattern to
that of the charge radii.

\subsection{Electric-quadrupole form factors}

   The results of the higher-order multipole moments must be regarded
as preliminary due to the lack of symmetry in the spatial dimensions
of our lattice.  The elongated $x$ dimension coupled with possible
finite volume restrictions may induce deformations from spherical
symmetry.  On the other hand, it is useful to examine these quantities
with a view toward determining the feasibility of extracting $E2$
moments in future calculations.

   The $E2$ form factor is particularly interesting since it provides
a glimpse into the shape of the baryon ground state.  A small $E2$
form factor would cast serious doubt on Skyrme models where the
hedgehog Skyrmion has an inherently large quadrupole moment.  Our
focus here is on estimating an upper bound for the $E2$ form factor.
Tables \ref{baryone2ff} and \ref{quarke2ff} summarize the lattice
results for $E2$ form factors.

   The negative sign of the central values for positively charged
baryons is consistent with deformation one might expect from our
elongated lattice.  Equation (\ref{ge2}) isolates the spin matrix
element $\bigm < {3 \over 2} s' \bigm | \Sigma^{(2)}_0 \bigm | {3
\over 2} s \bigm >$ and therefore determines the asymmetry of the
ground state wave-function having some overlap with the spherical
harmonic $Y_{20}(\theta)$ where $\theta$ is measured relative to the
spin quantization axis $\widehat z$.  A negative $E2$ moment
corresponds to an oblate shape which may be due to some broadening of
the wave function in the longer $x$ direction where overlap with
periodic images is minimized.

   In nonrelativistic models where angular momentum $l$ and spin $s$
are constants of the motion, it is useful to consider angular momentum
selection rules.  Consider the transition
\begin{equation}
\bigm < j'= l' + s' \bigm | j_\gamma \bigm | j = l + s \bigm > ,
\end{equation}
where $j_\gamma$ is the total angular momentum of the photon and
determines the multipole character of the transition.  Total angular
momentum selection rules require $j+j_\gamma=j'$ and therefore
\begin{equation}
| j - j' | \leq j_\gamma \leq j + j' .
\label{totalj}
\end{equation}
Combined with parity conservation, (\ref{totalj}) indicates there are
at most 4 multipole form factors available in spin-3/2 to spin-3/2
transitions.  Electric form factors correspond to $s=s'$.  This yields
the orbital angular momentum selection rule
\begin{equation}
| l - l' | \leq j_\gamma \leq l + l' ,
\end{equation}
indicating the $E2$ form factor vanishes in nonrelativistic models
unless some configuration mixing is included in the baryon ground
state.

   The relationship between our definition of the $E2$ form factor
given in (\ref{mege}) and that usually calculated in models may be
easily established by writing (\ref{mege}) using the spherical
harmonics in the Breit frame with $s=s'=j=3/2$.  Noting that
\begin{equation}
\left [ \widehat q \times \widehat q \right ]^{(2)}_m = \sqrt{ 8 \pi
\over 15} \, Y_{2m}(\widehat q) ,
\end{equation}
equation (\ref{mege}) may be written as
\FL
\begin{eqnarray}
\lefteqn{ \bigm < {\vec q \over 2}, {3 \over 2} \bigm | j^{0}(0)
\bigm | - {\vec q \over 2}, {3 \over 2} \bigm > = }
\qquad \qquad \nonumber \\
&& G_{E0}(q^{2}) + {2 \over 3} \tau \sqrt{4 \pi \over 5}
\, Y_{20}(\widehat q) \, G_{E2}(q^{2}) .
\end{eqnarray}
Model approaches determine current matrix elements via the fourier
transform
\FL
\begin{equation}
\bigm < {\vec q \over 2}, {3 \over 2} \bigm | j^{0}(0)
\bigm | - {\vec q \over 2}, {3 \over 2} \bigm > =
\int d^3r \, e^{i \vec q \cdot \vec r} \,
\overline \psi(r) \, j^0(r) \,
\psi(r) .
\end{equation}
Using the small $|\vec q|$ spherical harmonic expansion
\begin{equation}
e^{i \vec q \cdot \vec r} = 4 \pi \sum_{l=0}^{\infty}
{i^l \over (2 l + 1)!!} \, q^l r^l \sum_{m=-l}^l Y_{lm}^*(\widehat r)
\, Y_{lm}(\widehat q) \, ,
\end{equation}
and choosing $\vec q = q \widehat z$, it is straight forward to show
\begin{equation}
{\cal G}_{E2}(0) = M_B^2 \int d^3r \, \overline \psi(r) \,
(3 z^2 - r^2) \, \psi(r) ,
\end{equation}
where $3 z^2 - r^2$ is the standard operator used for quadrupole
moments.  The factor $M_B^2$ indicates that ${\cal G}_{E2}(0)$ gives
the $E2$ form factor in units of $(e/M_B^2)$.

   A chiral quark model calculation \cite{cohen86pc} gives ${\cal
G}_{E2}(0) = -0.30$ fm${}^2$ for $\Delta^{++}$ which lies $2 \sigma$
outside our bound of $-0.03(13)$ fm${}^2$.  However it may be more
appropriate to compare the dimensionless ratio of the $E2$ form factor
and the squared charge radius.  The generic Skyrme model result is
\cite{cohen86pc,adkins83}
\begin{equation}
{{\cal G}_{E2} \over \bigm < r^2 \bigm >_V } = - {4 \over 25} I_3 \, ,
\end{equation}
where $\bigm < r^2 \bigm >_V$ is the isovector charge radius and $I_3$
is the 3-component of isospin.  The analogous lattice ratio is
\begin{equation}
{{\cal G}_{E2} \over \bigm < r^2 \bigm >_V } =
\left (I_3 + {1 \over 2} \right )
{{\cal G}_{E2}^{\Delta^+} \over \bigm < r^2 \bigm >^{\Delta^+} } \, .
\end{equation}
The isospin dependence of the lattice results is proportional to the
baryon charge which is not the case for the Skyrme model.  For
$\Delta^{++}$ the lattice ratio is $-0.08(30)$ which encompasses the
Skyrme model ratio of $-0.24$.  Future high statistics lattice
calculations on a cubic lattice should be able to confirm or reject
the hedgehog Skyrmion description of baryons.

\subsection{Magnetic-octupole form factors}

   The magnetic-octupole form factors are large for heavy quark masses
and appear to decrease as the quark mass becomes lighter.  In the
physical regime the uncertainties are sufficiently large to make the
$M3$ moments consistent with zero with the exceptions of $\Xi^{*-}$
and $\Omega^-$ hyperons.  Tables \ref{baryonm3ff} and \ref{quarkm3ff}
summarize the lattice results for decuplet baryons and their quark
contributions respectively.

   Angular momentum selection rules for $M3$ moments require $s + 1 =
s'$ limiting nonvanishing transitions to
\begin{equation}
\left | | l - l' | - 1 \right | \leq j_\gamma \leq l + l' + 1 .
\end{equation}
Once again $M3$ transitions require nonzero orbital angular momentum
admixtures in the ground state wave-function.  Connection to model
calculations of $M3$ moments may be made from our definition of the
$M3$ moment in (\ref{megm}) in a manner analogous to that for the $E2$
moment.  $M3$ moments in hedgehog models \cite{cohen86pc} are $1/N_c$
suppressed relative to $M1$ moments and cannot been determined using
conventional semi-classical methods.

   The dimensionless ratio of interest for $M3$ moments relates the
$M3$ form factor to the $M1$ form factor and the squared charge
radius.  For $\Omega^-$ the ratio is
\begin{equation}
{{\cal G}_{M3} \over {\cal G}_{M1} \bigm < r^2 \bigm > } = 1.4(4) \, ,
\end{equation}
indicating a relatively large $M3$ moment.  This ratio is much larger
than one might expect from our discussion of the $E2$ form factor and
angular momentum selection rules.  For $\Omega^-$ the ratio is
\begin{equation}
{{\cal G}_{E2} \over \bigm < r^2 \bigm > } = 0.02(3) \, .
\end{equation}
However, in fully relativistic calculations angular momentum and spin
on their own are no longer good quantum numbers.  Gluons may also
carry angular momentum allowing radical changes from standard
$SU(6)$-spin-flavor symmetry as seen in the magnetic properties of
octet baryons and evidenced in the EMC polarized muon-proton
scattering experiment \cite{emc88}.  Furthermore, lattice QCD is a
relativistic theory with particle creation and annihilation allowing
some overlap with mesonic dressings of baryons even in the quenched
approximation.

   This pattern of small electric effects and large magnetic effects
is reminiscent of the electromagnetic properties of octet baryons.
The lattice results show a spin dependence in the quark distributions
that accounts for the negative squared charge radius of the neutron by
{\it slightly} increasing the $d$-quark distribution radius relative
to the $u$-quark radius.  However, spin dependence in the magnetic
properties is huge.  For example, figure \ref{octqmagmom} shows that
the effective magnetic moment of a $u$ quark in the neutron is roughly
half that of the $d$-quark when normalized to unit charge.

\section{Overview and Outlook}

   We have presented a fully relativistic formalism for isolating and
extracting the four electromagnetic multipole form factors of spin-3/2
systems in lattice field theory.  Results of the first lattice QCD
analysis of $SU(3)$-flavor decuplet baryons have been systematically
examined to reveal new aspects of the underlying nonperturbative
quark-gluon dynamics.

   The $E0$ and $M1$ correlation functions for decuplet baryons show a
broad plateau region allowing a reliable extraction of the
electromagnetic form factors.  Statistical uncertainties are similar
to that seen in our octet baryon analysis.

   The qualitative mass dependence of decuplet baryon charge radii
follows the anticipated pattern produced by a more localized strange
quark distribution.  The electric form factors of neutral baryons are
dominated by the net charge of the light quarks as in the octet baryon
results.  Model calculations of the $\Omega^-$ charge radius vary
widely and the lattice results favor a calculation based on a
relativised quark model \cite{barik90} over the other bag and Skyrme
models considered.

   Closer examination of the baryon charge radii reveals a behavior in
the quark charge distribution radii consistent with a spin dependent
force having an inverse relationship with the quark mass.  This
behavior is what one expects from the hyperfine interaction term of
the one-gluon-exchange potential.  The spin dependent force
counteracts center-of-mass shifts and suppresses baryon dependence of
the quark charge distribution radii.

   The center-of-mass shifts that give rise to an enhancement of the
$u$-quark magnetic moment in the octet $\Xi^0$ relative to that in the
neutron are offset by this repulsive force in decuplet baryons.
Variation in the quark effective magnetic moments from baryon to
baryon is minimal for decuplet baryons.  The residual effect is
largely due to the baryon mass setting the scale at which quarks
contribute to the baryon magnetic moment.

   The symmetric role of quarks in decuplet baryons contrasts that in
octet baryons.  The role of a quark in an octet baryon correlation
function is different depending on whether the quark is singly or
doubly represented.  This difference, which cannot arise in the
decuplet case, causes large variations in the electromagnetic
properties of quarks in octet baryons .

   The lattice predictions of decuplet baryon magnetic moments agree
with the simplest of quark models to a remarkable extent.  This is in
sharp contrast to the octet baryon analysis and is due to an
approximate baryon independence of effective quark magnetic moments in
decuplet baryons.  The lattice prediction of the $\Delta^{++}/p$
magnetic moment ratio is large compared to a recent experimental
analysis \cite{bosshard91} but compares better with previous analyses.
Our $\Omega^-$ magnetic moment agrees with experiment \cite{diehl91}
and prefers the simple quark model result and a calculation based on
effective quark masses \cite{chao90} over the other four model
calculations considered.

   The $E2$ and $M3$ form factors depend on subtle differences in the
correlation function ratios and as a result have larger statistical
uncertainties.  The $M3$ form factors are finite at all the values of
$\kappa$ considered.  The $E2$ form factor is at the threshold of
confirming or rejecting the hedgehog Skyrmion description of baryons.
A future high statistics lattice calculation on a cubic lattice will
provide considerable insight.

   For electromagnetic form factors to provide a reliable test of QCD,
one must understand and eliminate systematic errors.  While the
results of chiral perturbation theory may be used to assess the
magnitude of possible systematic errors in the lattice extrapolation
procedure \cite{leinweber92c}, there can be no substitute to actual
calculations probing regimes of lighter quark mass.

   A calculation of electromagnetic properties in full QCD would also
assist in understanding systematic errors.  However to calculate in
full QCD, the contributions of disconnected quark loops as illustrated
in figure \ref{disql} must be understood first.  These loop
contributions may play a key role in removing the discrepancies
between lattice and experimental violations of magnetic moment sum
rules \cite{leinweber92a}.  Closed quark loop contributions may be
estimated even in the quenched approximation and present a major
challenge for future lattice QCD calculations.  Such a calculation
will provide insight into important questions such as the strangeness
content of the proton and the general role played by sea quarks in
hadrons.

   The examination of decuplet baryon structure through the lattice
field theory approach to QCD has given a new and more detailed
understanding of nonperturbative interactions.  The underlying theme
of the results presented here involves a cancellation of
spin-dependent forces and center-of-mass effects which results in
structure that is consistent with nondynamical models such as the
simplest quark model based on $SU(6)$-spin-flavor symmetry broken only
by the quark masses.

   A future investigation \cite{leinweber92d} will address $N \gamma
\to \Delta$ transitions where there have been significant experimental
efforts to measure the $M1$ and $E2$ transition moments.  This, of
course, has been accompanied by a plethora of model calculations for
these quantities.  The lattice results will be instrumental in
assessing the reliability of the model analyses.

   The odd-parity spin-3/2 baryon octet also offers an interesting
forum for checking model predictions.  There has not been a lattice
investigation of these states.  The $N(1520)$ is the lowest lying
baryon with $I(J^p)={1 \over 2}({3 \over 2}^-)$. It is expected to be
stable with our present lattice parameters and therefore is accessible
to lattice calculations.  The neighboring $N{1 \over 2}^-(1535)$ is a
source of possible contamination in the $j=3/2$ interpolating field
and therefore angular momentum projection operators may be required
\cite{leinweber90}.  Calculations of the electromagnetic properties of
$N{3 \over 2}^-(1520)$ may be completely different from model
expectations if gluons carry a significant fraction of the angular
momentum usually attributed to quark degrees of freedom alone.  With
the anticipated experimental program at CEBAF, a lattice QCD analysis
of $N{1 \over 2}^+ \gamma \to N{3 \over 2}^-$ transition moments is
also of interest.

   It is very encouraging that lattice QCD evaluations of hadronic
electromagnetic form factors give, for example, a pattern of magnetic
moments which is as good or better than hadronic models with
adjustable parameters which in some cases have been tuned to fit
magnetic moments.  With further study to understand and eliminate
systematic errors, we believe the study of hadron electromagnetic form
factors will prove ultimately to be one of the best quantitative
testing grounds for nonperturbative QCD.

\acknowledgements

   The computing resources for this study were provided by the
Computing Science Department and the Center for Computational Sciences
at the University of Kentucky on their IBM 3090-600J supercomputer.
D.B.L. thanks Thomas Cohen, Wojciech Broniowski and Manoj Banerjee for
enthusiastic discussions.  T.D. thanks Keh-Fei Liu for many useful
conversations.  This work is supported in part by the U.S. Department
of Energy under grant numbers DE-FG05-87ER-40322 and
DE-FG05-84ER-40154, the National Science Foundation under grant number
STI-9108764 and the Natural Sciences and Engineering Research Council
of Canada.


\begin{thebibliography}{10}
\bibitem{correlations}
This proved sufficient to achieve good statistical independence of the
ensemble. An examination of a number of observables, configuration by
configuration, revealed no apparent long-range correlation in our
gauge configuration sample.
\bibitem{raddiff}
While there is some overlap in the error bars of figure
\ref{octsigqr}, a calculation of the change of the strange quark
distribution radius indicates the radius is strictly decreasing as the
$u$ quarks become lighter. This is not the case for the decuplet
$\Sigma^{*+}$.
\end{thebibliography}

\newpage

\mediumtext
\begin{table}
\caption{Baryon Mass in lattice units $(M_B a)$.}
\label{baryonmass}
\setdec 0.00(0)
\begin{tabular}{lcccc}
Baryon &$\kappa_1=0.152$ &$\kappa_2=0.154$ &$\kappa_3=0.156$
       &$\kappa_{cr}=0.159\,8(2)$\\
\tableline
$N       $ &\dec 1.09(3) &\dec 0.96(3) &\dec 0.84(3) &\dec 0.61(5)  \\
$\Delta  $ &\dec 1.13(3) &\dec 1.02(4) &\dec 0.90(5) &\dec 0.70(7)  \\
$\Sigma^*$ &\dec 1.13(3) &\dec 1.05(4) &\dec 0.98(4) &\dec 0.84(5)  \\
$\Xi^*   $ &\dec 1.13(3) &\dec 1.09(4) &\dec 1.05(4) &\dec 0.98(4)  \\
$\Omega  $ &\dec 1.13(3) &             &             &\dec 1.13(3)  \\
\end{tabular}
\end{table}

\begin{table}
\caption{Baryon electric charge form factors.}
\label{baryone0ff}
\setdec 00.000(00)
\begin{tabular}{lcccc}
Baryon &$\kappa_1=0.152$ &$\kappa_2=0.154$ &$\kappa_3=0.156$
       &$\kappa_{cr}=0.159\,8(2)$\\
\tableline
$\Delta^{++}$ &\dec 1.723(17)    &\dec 1.681(27)
              &\dec 1.636(55)    &\dec 1.562(80)    \\
$\Delta^{+} $ &\dec 0.861(9)     &\dec 0.841(13)
              &\dec 0.818(27)    &\dec 0.781(40)    \\
$\Delta^{0} $ &\dec 0.000        &\dec 0.000
              &\dec 0.000        &\dec 0.000        \\
$\Delta^{-} $ &\dec $-$0.861(9)  &\dec $-$0.841(13)
              &\dec $-$0.818(27) &\dec $-$0.781(40) \\
$\Sigma^{*+}$ &\dec 0.861(9)     &\dec 0.831(13)
              &\dec 0.796(28)    &\dec 0.743(40)    \\
$\Sigma^{*0}$ &\dec 0.000        &\dec $-$0.008(1)
              &\dec $-$0.017(5)  &\dec $-$0.032(8)  \\
$\Sigma^{*-}$ &\dec $-$0.861(9)  &\dec $-$0.847(11)
              &\dec $-$0.831(19) &\dec $-$0.805(28) \\
$\Xi^{*0}   $ &\dec 0.000        &\dec $-$0.016(3)
              &\dec $-$0.036(8)  &\dec $-$0.065(14) \\
$\Xi^{*-}   $ &\dec $-$0.861(9)  &\dec $-$0.854(10)
              &\dec $-$0.846(12) &\dec $-$0.832(16) \\
$\Omega^{-} $ &\dec $-$0.861(9)  &
              &                  &\dec $-$0.861(9)  \\
\end{tabular}
\end{table}

\begin{table}
\caption{Electric charge form factors for single quarks of unit
         charge.}
\label{quarke0ff}
\setdec 00.000(00)
\begin{tabular}{lcccc}
Quark &$\kappa_1=0.152$ &$\kappa_2=0.154$ &$\kappa_3=0.156$
&$\kappa_{cr}=0.159\,8(2)$ \\
\tableline
$u_\Delta    $ &\dec 0.861(9)  &\dec 0.841(13)
               &\dec 0.818(27) &\dec 0.781(40) \\
$u_{\Sigma^*}$ &\dec 0.861(9)  &\dec 0.839(12)
               &\dec 0.814(24) &\dec 0.774(34) \\
$s_{\Sigma^*}$ &\dec 0.861(9)  &\dec 0.863(9)
               &\dec 0.866(13) &\dec 0.869(18) \\
$u_{\Xi^*}   $ &\dec 0.861(9)  &\dec 0.838(12)
               &\dec 0.810(19) &\dec 0.767(27) \\
$s_{\Xi^*}   $ &\dec 0.861(9)  &\dec 0.862(9)
               &\dec 0.863(9)  &\dec 0.865(11) \\
$s_\Omega    $ &\dec 0.861(9)  &
               &               &\dec 0.861(9)  \\
\end{tabular}
\end{table}

\begin{table}
\caption{Baryon rms charge radii normalized to unit charge in
         lattice units \break $\bigm < r^2/a^2 \bigm >^{1/2}$.}
\label{baryonchgrad}
\setdec 0.00(00)
\begin{tabular}{lcccc}
Baryon &$\kappa_1=0.152$ &$\kappa_2=0.154$ &$\kappa_3=0.156$
       &$\kappa_{cr}=0.159\,8(2)$\\
\tableline
$\Delta^{++} $ &\dec 3.71(13) &\dec 4.02(19)
               &\dec 4.34(39) &\dec 4.90(57) \\
$\Delta^{+}  $ &\dec 3.71(13) &\dec 4.02(19)
               &\dec 4.34(39) &\dec 4.90(57) \\
$\Delta^{-}  $ &\dec 3.71(13) &\dec 4.02(19)
               &\dec 4.34(39) &\dec 4.90(57) \\
$\Sigma^{*+} $ &\dec 3.71(13) &\dec 4.15(19)
               &\dec 4.63(38) &\dec 5.42(56) \\
$\Sigma^{*-} $ &\dec 3.71(13) &\dec 3.92(16)
               &\dec 4.16(28) &\dec 4.54(41) \\
$\Xi^{*-}    $ &\dec 3.71(13) &\dec 3.81(14)
               &\dec 3.94(18) &\dec 4.15(23) \\
$\Omega^{-}  $ &\dec 3.71(13) &
               &              &\dec 3.71(13) \\
\end{tabular}
\end{table}

\begin{table}
\caption{Rms charge radii for single quarks of unit charge in
         lattice units $\bigm < r^2/a^2 \bigm >^{1/2}$.}
\label{quarkchgrad}
\setdec 0.00(00)
\begin{tabular}{lcccc}
Quark &$\kappa_1=0.152$ &$\kappa_2=0.154$ &$\kappa_3=0.156$
      &$\kappa_{cr}=0.159\,8(2)$ \\
\tableline
$u_\Delta    $ &\dec 3.71(13)  &\dec 4.02(19)
               &\dec 4.34(39)  &\dec 4.90(57)  \\
$u_{\Sigma^*}$ &\dec 3.71(13)  &\dec 4.03(18)
               &\dec 4.40(33)  &\dec 4.99(48)  \\
$s_{\Sigma^*}$ &\dec 3.71(13)  &\dec 3.68(14)
               &\dec 3.65(20)  &\dec 3.60(28)  \\
$u_{\Xi^*}   $ &\dec 3.71(13)  &\dec 4.05(17)
               &\dec 4.44(27)  &\dec 5.07(38)  \\
$s_{\Xi^*}   $ &\dec 3.71(13)  &\dec 3.70(13)
               &\dec 3.68(14)  &\dec 3.66(16)  \\
$s_\Omega    $ &\dec 3.71(13)  &
               &               &\dec 3.71(13)  \\
\end{tabular}
\end{table}

\narrowtext
\begin{table}
\caption{Neutral baryon rms charge radii in units of the proton radius
         $\bigm < |r^2/r_p^2| \bigm >^{1/2}$.}
\label{neutchgrad}
\setdec 0.00(00)
\begin{tabular}{lc}
Baryon &Radius \\
\tableline
$n$           &\dec 0.34(22) \\
$\Lambda^{0}$ &\dec 0.39(8)  \\
$\Sigma^{0} $ &\dec 0.44(4)  \\
$\Xi^{0}    $ &\dec 0.53(10) \\
$\Sigma^{*0}$ &\dec 0.39(4)  \\
$\Xi^{*0}   $ &\dec 0.57(9)  \\
\end{tabular}
\end{table}

\mediumtext
\begin{table}
\caption{Baryon magnetic-dipole $(M1)$ form factors in units of
         natural magnetons $(\mu_B = e/2M_B)$.}
\label{baryonm1ff}
\setdec 00.000(00)
\begin{tabular}{lcccc}
Baryon &$\kappa_1=0.152$ &$\kappa_2=0.154$ &$\kappa_3=0.156$
       &$\kappa_{cr}=0.159\,8(2)$\\
\tableline
$\Delta^{++} $ &\dec 4.48(30)    &\dec 4.48(42)
               &\dec 4.45(60)    &\dec 4.44(87)  \\
$\Delta^{+}  $ &\dec 2.24(15)    &\dec 2.24(21)
               &\dec 2.22(30)    &\dec 2.22(44)  \\
$\Delta^{0}  $ &\dec 0.00        &\dec 0.00
               &\dec 0.00        &\dec 0.00     \\
$\Delta^{-}  $ &\dec $-$2.24(15) &\dec $-$2.24(21)
               &\dec $-$2.22(30) &\dec $-$2.22(44) \\
$\Sigma^{*+} $ &\dec 2.24(15)    &\dec 2.36(21)
               &\dec 2.47(30)    &\dec 2.69(44)  \\
$\Sigma^{*0} $ &\dec 0.00        &\dec 0.058(13)
               &\dec 0.119(45)   &\dec 0.224(75) \\
$\Sigma^{*-} $ &\dec $-$2.24(15) &\dec $-$2.25(19)
               &\dec $-$2.24(24) &\dec $-$2.24(32) \\
$\Xi^{*0}    $ &\dec 0.00        &\dec 0.119(23)
               &\dec 0.235(68)   &\dec 0.45(12)  \\
$\Xi^{*-}    $ &\dec $-$2.24(15) &\dec $-$2.25(17)
               &\dec $-$2.24(19) &\dec $-$2.24(23) \\
$\Omega^{-}  $ &\dec $-$2.24(15) &
               &                 &\dec $-$2.24(15) \\
\end{tabular}
\end{table}

\begin{table}
\caption{
Magnetic-dipole $(M1)$ form factors for single quarks of unit charge
in units of natural magnetons $(\mu_B = e/2M_B)$.}
\label{quarkm1ff}
\setdec 00.000(00)
\begin{tabular}{lcccc}
Quark &$\kappa_1=0.152$ &$\kappa_2=0.154$ &$\kappa_3=0.156$
      &$\kappa_{cr}=0.159\,8(2)$ \\
\tableline
$u_\Delta    $ &\dec 2.24(15)  &\dec 2.24(21)
               &\dec 2.22(30)  &\dec 2.22(44)  \\
$u_{\Sigma^*}$ &\dec 2.24(15)  &\dec 2.30(20)
               &\dec 2.36(27)  &\dec 2.46(38)  \\
$s_{\Sigma^*}$ &\dec 2.24(15)  &\dec 2.13(17)
               &\dec 2.00(20)  &\dec 1.79(25)  \\
$u_{\Xi^*}   $ &\dec 2.24(15)  &\dec 2.36(19)
               &\dec 2.48(24)  &\dec 2.69(32)  \\
$s_{\Xi^*}   $ &\dec 2.24(15)  &\dec 2.19(16)
               &\dec 2.12(17)  &\dec 2.02(19)  \\
$s_\Omega    $ &\dec 2.24(15)  &
               &               &\dec 2.24(15)  \\
\end{tabular}
\end{table}

\begin{table}
\caption{
Lattice results for decuplet baryon magnetic moments in units of
nuclear magnetons $(\mu_N)$, ratios of decuplet baryon moments to the
proton magnetic moment and ratios scaled to reproduce the proton
magnetic moment.}
\label{baryonmagmom}
\setdec 00.000(00)
\begin{tabular}{lccc}
Baryon &Magnetic moment &Ratio $(\mu/\mu_p)$ &Scaled \\
       &$(\mu_N)$       &                    &$(\mu_N)$ \\
\tableline
$\Delta^{++}$ &\dec     4.91(61)
              &\dec     2.18(32)  &\dec     6.09(88)  \\
$\Delta^{+} $ &\dec     2.46(31)
              &\dec     1.09(16)  &\dec     3.05(44)  \\
$\Delta^{0} $ &\dec     0.00
              &\dec     0.00      &\dec     0.00      \\
$\Delta^{-} $ &\dec  $-$2.46(31)
              &\dec  $-$1.09(16)  &\dec  $-$3.05(44)  \\
$\Sigma^{*+}$ &\dec     2.55(26)
              &\dec     1.13(14)  &\dec     3.16(40)  \\
$\Sigma^{*0}$ &\dec     0.27(5)
              &\dec     0.118(24) &\dec     0.329(67) \\
$\Sigma^{*-}$ &\dec  $-$2.02(18)
              &\dec  $-$0.90(10)  &\dec  $-$2.50(29)  \\
$\Xi^{*0}   $ &\dec     0.46(7)
              &\dec     0.206(35) &\dec     0.58(10)  \\
$\Xi^{*-}   $ &\dec  $-$1.68(12)
              &\dec  $-$0.744(85) &\dec  $-$2.08(24)  \\
$\Omega^{-} $ &\dec  $-$1.40(10)
              &\dec  $-$0.621(78) &\dec  $-$1.73(22)  \\
\end{tabular}
\end{table}

\narrowtext
\begin{table}
\caption{
Effective quark magnetic moments for quarks of unit charge in units of
nuclear magnetons $(\mu_N)$.}
\label{quarkeffmom}
\setdec 0.00(00)
\begin{tabular}{lclc}
\multicolumn{2}{c}{Octet Baryons}
&\multicolumn{2}{c}{Decuplet Baryons} \\
Quark &Effective moment &Quark &Effective moment \\
\tableline
$u_p$          &\dec 2.41(27)  &$u_\Delta$     &\dec 2.46(31)  \\
$d_p$          &\dec 1.10(74)  &$d_\Delta$     &\dec 2.46(31)  \\
$u_\Sigma$     &\dec 2.12(23)  &$u_{\Sigma^*}$ &\dec 2.29(22)  \\
$s_\Sigma$     &\dec 0.59(37)  &$s_{\Sigma^*}$ &\dec 1.49(13)  \\
$u_\Xi   $     &\dec 1.53(27)  &$u_{\Xi^*}   $ &\dec 2.14(18)  \\
$s_\Xi   $     &\dec 1.24(10)  &$s_{\Xi^*}   $ &\dec 1.45(10)  \\
$s_\Lambda$    &\dec 1.25(15)  &$s_\Omega    $ &\dec 1.40(10)  \\
\end{tabular}
\end{table}

\begin{table}
\caption{Effective $u$- and $s$-quark magnetic moment ratios.}
\label{eqmmr}
\begin{tabular}{lc}
Quark Ratio &Lattice Result \\
\tableline
$u_{\Sigma^*}/u_{\Delta}$  &0.93(5) \\
$u_{\Xi^*}/u_{\Sigma^*}$   &0.94(3) \\
$u_{\Xi^*}/u_{\Delta}$     &0.87(7) \\
$s_{\Xi^*}/s_{\Sigma^*}$   &0.97(5) \\
$s_{\Omega}/s_{\Xi^*}$     &0.97(3) \\
$s_{\Omega}/s_{\Sigma^*}$  &0.94(7) \\
\end{tabular}
\end{table}

\mediumtext
\begin{table}
\caption{Baryon Magnetic Radii
 $\sqrt{\bigm < r^2 \bigm > / \left ( {\cal G}_{M1} a^2 \right ) }$
         in lattice units.}
\label{baryonmagrad}
\setdec 00.000(00)
\begin{tabular}{lcccc}
Baryon &$\kappa_1=0.152$ &$\kappa_2=0.154$ &$\kappa_3=0.156$
       &$\kappa_{cr}=0.159\,8(2)$\\
\tableline
$\Delta^{++}$ &\dec 3.71(13)  &\dec 4.02(19)
              &\dec 4.34(39)  &\dec  4.90(57) \\
$\Delta^{+} $ &\dec 3.71(13)  &\dec 4.02(19)
              &\dec 4.34(39)  &\dec  4.90(57) \\
$\Delta^{-} $ &\dec 3.71(13)  &\dec 4.02(19)
              &\dec 4.34(39)  &\dec  4.90(57) \\
$\Sigma^{*+}$ &\dec 3.71(13)  &\dec 4.13(19)
              &\dec 4.58(38)  &\dec  5.34(56) \\
$\Sigma^{*-}$ &\dec 3.71(13)  &\dec 3.93(17)
              &\dec 4.19(29)  &\dec  4.59(41) \\
$\Xi^{*-}   $ &\dec 3.71(13)  &\dec 3.82(14)
              &\dec 3.98(18)  &\dec  4.22(23) \\
$\Omega^{-} $ &\dec 3.71(13)  &
              &               &\dec  3.71(13) \\
\end{tabular}
\end{table}

\begin{table}
\caption{Baryon electric-quadrupole $(E2)$ form factors in units
         of $(e/M_B^2)$.}
\label{baryone2ff}
\setdec 00.00(00)
\begin{tabular}{lcccc}
Baryon &$\kappa_1=0.152$ &$\kappa_2=0.154$ &$\kappa_3=0.156$
       &$\kappa_{cr}=0.159\,8(2)$\\
\tableline
$\Delta^{++}$ &\dec $-$0.6(8)  &\dec $-$0.8(14)
              &\dec $-$0.7(28) &\dec $-$1.0(39)  \\
$\Delta^{+} $ &\dec $-$0.3(4)  &\dec $-$0.4(7)
              &\dec $-$0.4(14) &\dec $-$0.5(19)  \\
$\Delta^{0} $ &\dec 0.0        &\dec 0.0
              &\dec 0.0        &\dec  0.0      \\
$\Delta^{-} $ &\dec 0.3(4)     &\dec 0.4(7)
              &\dec 0.4(14)    &\dec  0.5(19)  \\
$\Sigma^{*+}$ &\dec $-$0.3(4)  &\dec $-$0.3(7)
              &\dec $-$0.1(13) &\dec $-$0.1(19)  \\
$\Sigma^{*0}$ &\dec 0.00       &\dec 0.05(7)
              &\dec 0.15(23)   &\dec  0.26(36) \\
$\Sigma^{*-}$ &\dec 0.3(4)     &\dec 0.4(6)
              &\dec 0.4(9)     &\dec  0.5(13)  \\
$\Xi^{*0}   $ &\dec 0.00       &\dec 0.07(12)
              &\dec 0.15(31)   &\dec  0.29(54) \\
$\Xi^{*-}   $ &\dec 0.3(4)     &\dec 0.3(5)
              &\dec 0.4(6)     &\dec  0.4(8)   \\
$\Omega^{-} $ &\dec 0.3(4)     &
              &                &\dec  0.3(4)   \\
\end{tabular}
\end{table}

\begin{table}
\caption{Electric-quadrupole $(E2)$ form factors for single quarks
         of unit charge in units of $(e/M_B^2)$.}
\label{quarke2ff}
\setdec 00.0(00)
\begin{tabular}{lcccc}
Quark &$\kappa_1=0.152$ &$\kappa_2=0.154$ &$\kappa_3=0.156$
      &$\kappa_{cr}=0.159\,8(2)$ \\
\tableline
$u_\Delta    $ &\dec $-$0.3(4)  &\dec $-$0.4(7)
               &\dec $-$0.4(14) &\dec $-$0.5(19) \\
$u_{\Sigma^*}$ &\dec $-$0.3(4)  &\dec $-$0.3(6)
               &\dec $-$0.3(11) &\dec $-$0.3(16) \\
$s_{\Sigma^*}$ &\dec $-$0.3(4)  &\dec $-$0.4(5)
               &\dec $-$0.7(6)  &\dec $-$1.0(8)  \\
$u_{\Xi^*}   $ &\dec $-$0.3(4)  &\dec $-$0.3(6)
               &\dec $-$0.2(9)  &\dec $-$0.2(12) \\
$s_{\Xi^*}   $ &\dec $-$0.3(4)  &\dec $-$0.4(4)
               &\dec $-$0.5(5)  &\dec $-$0.6(6)  \\
$s_\Omega    $ &\dec $-$0.3(4)  &
               &                &\dec $-$0.3(4)  \\
\end{tabular}
\end{table}

\begin{table}
\caption{Baryon magnetic-octupole $(M3)$ form factors in units
         of $(e/2 M_B^3)$.}
\label{baryonm3ff}
\begin{tabular}{lr@{$\pm$}lr@{$\pm$}lr@{$\pm$}lr@{$\pm$}l}
Baryon &\multicolumn{2}{c}{$\kappa_1=0.152$}
       &\multicolumn{2}{c}{$\kappa_2=0.154$}
       &\multicolumn{2}{c}{$\kappa_3=0.156$}
       &\multicolumn{2}{c}{$\kappa_{cr}=0.159\,8(2)$} \\
\tableline
$\Delta^{++}$ &    113 &30   &     99 &51
              &     80 &73   &     55 &112  \\
$\Delta^{+} $ &     56 &15   &     50 &25
              &     40 &37   &     27 &56   \\
$\Delta^{0} $ &      0 &0    &      0 &0
              &      0 &0    &      0 &0    \\
$\Delta^{-} $ &  $-$56 &15   &  $-$50 &25
              &  $-$40 &37   &  $-$27 &56   \\
$\Sigma^{*+}$ &     56 &15   &     54 &27
              &     48 &48   &     43 &72   \\
$\Sigma^{*0}$ &      0 &0    &    0.8 &2.9
              & $-$0.4 &7.7  &    0.3 &1.4  \\
$\Sigma^{*-}$ &  $-$56 &15   &  $-$52 &22
              &  $-$49 &34   &  $-$41 &49   \\
$\Xi^{*0}   $ &      0 &0    &      2 &5
              &      2 &14   &      6 &24   \\
$\Xi^{*-}   $ &  $-$56 &15   &  $-$54 &19
              &  $-$54 &25   &  $-$51 &32   \\
$\Omega^{-} $ &  $-$56 &15   &  $-$56 &15
              &  $-$56 &15   &  $-$56 &15   \\
\end{tabular}
\end{table}

\begin{table}
\caption{Magnetic-octupole $(M3)$ form factors for single quarks
         of unit charge in units of $(e/2 M_B^3)$.}
\label{quarkm3ff}
\begin{tabular}{lr@{$\pm$}lr@{$\pm$}lr@{$\pm$}lr@{$\pm$}l}
Quark  &\multicolumn{2}{c}{$\kappa_1=0.152$}
       &\multicolumn{2}{c}{$\kappa_2=0.154$}
       &\multicolumn{2}{c}{$\kappa_3=0.156$}
       &\multicolumn{2}{c}{$\kappa_{cr}=0.159\,8(2)$} \\
\tableline
$u_\Delta    $ &   56 &15  &   50 &25   &   40 &37   &   27 &56   \\
$u_{\Sigma^*}$ &   56 &15  &   53 &24   &   48 &41   &   42 &61   \\
$s_{\Sigma^*}$ &   56 &15  &   51 &17   &   49 &22   &   41 &28   \\
$u_{\Xi^*}   $ &   56 &15  &   57 &23   &   56 &37   &   57 &55   \\
$s_{\Xi^*}   $ &   56 &15  &   53 &16   &   52 &19   &   48 &22   \\
$s_\Omega    $ &   56 &15  &   56 &15   &   56 &15   &   56 &15   \\
\end{tabular}
\end{table}

\narrowtext
\figure{
Octet and decuplet baryon masses.  The nucleon mass has been used to
set the scale.  Horizontal lines indicate the experimental values.
\label{decmass}}

\figure{
Quark contributions to the correlation function ratios
isolating the $\Sigma^{*+}$ electric-charge form factor as in
(\ref{ge0}).  The correlation functions are for $\kappa_u = \kappa_3$.
Ratios for larger quark mass have smaller statistical uncertainties.
\label{sige0R3}}

\figure{
Quark contributions to the correlation function ratios
isolating the $\Xi^{*0}$ magnetic-dipole form factor as in
(\ref{gm1}).  For purposes of illustration the leading kinematical
factor, which varies in the second order jackknife procedure, has been
set to a constant.  The correlation functions are for $\kappa_u =
\kappa_3$.  Ratios for larger quark mass have smaller statistical
uncertainties.
\label{xim1R3}}

\figure{
A comparison of the ratio sums in the parentheses of equations
(\ref{ge0}) and (\ref{ge2}) isolating ${\cal G}_{E0}$ and ${\cal
G}_{E2}$ respectively.  Correlation function ratios are for $\kappa_u
= \kappa_1$.
\label{dele2e0R3}}

\figure{
A comparison of the ratio sums in the parentheses of equations
(\ref{gm1}) and (\ref{gm3}) isolating ${\cal G}_{M1}$ and ${\cal
G}_{M3}$ respectively.  The $M3$ form factor is finite.  However, the
ratio fails to form a clear plateau.  Correlation function ratios are
for $\kappa_u = \kappa_1$.
\label{delm3m1R3}}

\figure{
Electric charge radii for charged octet and decuplet baryons in units
of the proton charge radius.  Radii of $\Delta^{++}$ and $\Delta^-$
are equal to that of $\Delta^+$ when normalized by the total baryon
charge.
\label{chgrad}}

\figure{
Extrapolations of the electric charge distributions of quarks within
the octet $\Sigma^+$ baryon.  The radius of the $s$-quark distribution
decreases as the $u$ quarks become lighter.
\label{octsigqr}}

\figure{
Extrapolations of the electric charge distributions of quarks within
the decuplet $\Sigma^{*+}$ baryon.  The radius of the $s$-quark
distribution is largely unaffected by changes in the $u$-quark mass.
\label{decsigqr}}

\figure{
Charge distribution radii of unit charge quarks within decuplet
baryons.  Symmetric isospin symmetry equates $u$- and $d$-quark
properties in decuplet baryons.  No significant baryon dependence of
the charge radii is seen.
\label{qrkradii}}

\figure{
Charge distribution radii of unit charge quarks within octet baryons.
Center-of-mass shifts and spin-dependent forces give rise to
significant baryon dependence of the charge radii.
\label{oqchgrad}}

\figure{
Ratios of octet and decuplet quark charge distribution radii.  Doubly
represented quarks show little octet/decuplet baryon dependence.
\label{qchgradrat}}

\figure{
The $\Omega^-/p$ charge radius ratio.  The model results include a
calculation based on a relativised quark model (Q.M.), MIT bag (MIT)
and cloudy bag (C.B.) models, and a bound state approach Skyrme
(Skyr.) model.  References may be found in the text.
\label{omegarad}}

\figure{
Electric charge form factors of neutral baryons calculated at $q
\simeq 0.4$ GeV.  The light quark charges dominate the hyperon form
factors while decuplet spin symmetry causes the $\Delta^0$ form factor
to vanish.
\label{e0neut}}

\figure{
Ratios of the lattice baryon magnetic moments to the lattice proton
result scaled to reproduce the proton magnetic moment.  Horizontal
dashed lines indicate the simple quark model predictions.
\label{decmagmqm}}

\figure{
Effective quark moments in decuplet baryons.  Approximate baryon
independence of the quark moments is displayed.
\label{qrkmagm}}

\figure{
Effective quark moments in octet baryons.  Significant baryon
dependence of the quark moments is illustrated.
\label{octqmagmom}}

\figure{
Ratios of octet baryon effective quark moments with their decuplet
partners.  Octet/decuplet dependence is seen throughout the quarks
with the exception of $u_p/u_\Delta$ and possibly
$u_\Sigma/u_{\Sigma^*}$.
\label{qmagmomrat}}

\figure{
Comparison of the lattice $\Delta^{++}/p$ magnetic moment ratio
(Latt.) with model calculations and an experimental result (Expt.)
having some model dependence.  The model calculations are the simple
quark model (Q.M.), cloudy bag (C.B.), Skyrme (Skyr.), Bethe-Salpeter
(B.S.) and QCD sum rules (S.R.).  References are given in the text.
\label{deltapp}}

\figure{
Comparison of the lattice $\Omega^-/p$ magnetic moment ratio (Latt.)
with model calculations and experiment (Expt.).  The model
calculations are as in figure \ref{deltapp} with additions of an
additive quark model based on effective quark masses (E.M.) and a
calculation including relativistic corrections (R.C.).  References are
given in the text.
\label{omegamin}}

\figure{
Magnetic radii of decuplet baryons in units of the proton magnetic
radius.  The magnetic radii follow a similar pattern to that of the
charge radii.
\label{dbmagrad}}

\figure{A skeleton diagram of a disconnected quark loop contributing
to the magnetic moment of a baryon.  The diagram may be dressed with
an arbitrary number of gluons.
\label{disql}}

\end{document}